\documentclass[reqno,a4paper,final,11pt]{amsart}
\usepackage[utf8x]{inputenc}
\usepackage{amssymb,amstext,amsthm,amscd,eucal}
\usepackage{color}
\usepackage{a4wide}
\usepackage{array}
\usepackage[all]{xy}
\usepackage{hyperref}
\hypersetup{%
  pdftitle   = {Killing superalgebra deformations of ten-dimensional
    supergravity backgrounds},
  pdfkeywords = {M-theory, supergravity, supersymmetry, superalgebra,
    deformations}
  pdfauthor  = {José Figueroa-O'Farrill and Bert Vercnocke},
  pdfcreator = {\LaTeX\ with package \flqq hyperref\frqq}
}
\PrerenderUnicode{éÉ}
\newcommand{\half}{\tfrac12}
\newcommand{\ff}{\mathfrak{f}}
\newcommand{\fg}{\mathfrak{g}}

\newcommand{\fk}{\mathfrak{k}}
\newcommand{\fh}{\mathfrak{h}}
\newcommand{\fp}{\mathfrak{p}}
\newcommand{\fn}{\mathfrak{n}}
\newcommand{\fr}{\mathfrak{r}}
\newcommand{\fs}{\mathfrak{s}}

\newcommand{\fso}{\mathfrak{so}}

\newcommand{\fosp}{\mathfrak{osp}}
\newcommand{\fsu}{\mathfrak{su}}

\newcommand{\Cl}{\mathrm{C}\ell}

\newcommand{\ten}{\natural}
\ifx\Sp\UnDeFiNeD
\newcommand{\Sp}{\mathrm{Sp}}
\else
\renewcommand{\Sp}{\mathrm{Sp}}
\fi

\newcommand{\NS}{\mathrm{NS}}
\newcommand{\D}{\mathrm{D}}
\newcommand{\DM}{\mathrm{dM}}
\newcommand{\F}{\mathrm{F}}

\newcommand{\RR}{\mathbb{R}}

\newcommand{\CC}{\mathbb{C}}

\renewcommand{\1}{\boldsymbol{1}}

\newcommand{\be}{\boldsymbol{e}}
\newcommand{\beps}{\boldsymbol{\varepsilon}}
\newcommand{\bpsi}{\boldsymbol{\psi}}
\newcommand{\bnu}{\boldsymbol{\nu}}

\newcommand{\alphadot}{{\dot\alpha}}
\newcommand{\betadot}{{\dot\beta}}
\newcommand{\gammadot}{{\dot\gamma}}
\newcommand{\alphabar}{{\bar\alpha}}
\newcommand{\betabar}{{\bar\beta}}
\newcommand{\gammabar}{{\bar\gamma}}
\DeclareMathOperator{\AdS}{AdS}

\DeclareMathOperator{\End}{End}

\DeclareMathOperator{\Hom}{Hom}

\newcommand{\MUNCH}[1]{\relax}

\allowdisplaybreaks[1]
\begin{document}
\title[Deformations of ten-dimensional Killing superalgebras]{Killing
  superalgebra deformations of ten-dimensional supergravity backgrounds}
\author[Figueroa-O'Farrill]{José Figueroa-O'Farrill}
\address{Maxwell Institute and School of Mathematics, The University
  of Edinburgh}
\email{J.M.Figueroa@ed.ac.uk}
\author[Vercnocke]{Bert Vercnocke}
\address{Instituut voor Theoretische Fysica, Katholieke Universiteit Leuven}
\email{Bert.Vercnocke@fys.kuleuven.be}
\thanks{EMPG-07-14}
%\date{\today}
\begin{abstract}
  We explore Lie superalgebra deformations of the Killing
  superalgebras of some ten-dimensional supergravity backgrounds.  We
  prove the rigidity of the Poincaré superalgebras in types I, IIA and
  IIB, as well as of the Killing superalgebra of the Freund--Rubin
  vacuum of type IIB supergravity.  We also prove rigidity of the
  Killing superalgebras of the NS5, D0, D3, D4 and D5 branes, whereas
  we exhibit the possible deformations of the D1, D2, D6 and D7 brane
  Killing superalgebras, as well as of that of the type II fundamental
  string solutions.  We relate the superalgebra deformations of the D2
  and D6 branes to those of the (delocalised) M2 brane and the
  Kaluza--Klein monopole, respectively.  The good behaviour under
  Kaluza--Klein reduction suggests that the deformed superalgebras
  ought to have a geometric interpretation.
\end{abstract}
\maketitle
\tableofcontents

\section{Introduction}
\label{sec:intro}

This paper continues the study initiated in \cite{JMFSuperDeform} of
Lie superalgebra deformations of Killing superalgebras of supergravity
backgrounds \cite{AFHS, GMT1, GMT2, PKT, JMFKilling, FOPFlux, NewIIB,
  ALOKilling, FMPHom, EHJGMHom}.  The focus in \cite{JMFSuperDeform}
was on eleven-dimensional supergravity backgrounds: the Minkowski and
Freund--Rubin backgrounds, whose Killing superalgebras were shown to
be rigid, as well as the elementary branes and the elementary purely
gravitational backgrounds.  Of these, the M5-brane Killing
superalgebra is rigid, but all the others admit deformations.  The
physical interpretation, if any, of these deformations was not
explored in \cite{JMFSuperDeform}: they could be due perhaps to
quantum corrections or perhaps to geometric limits within classical
supergravity.  This latter possibility is explored further in
\cite{FigRitDef}; although in the present paper we shall give indirect
evidence for the geometric origin of such deformations.  In this paper
we treat the case of types I and II ten-dimensional supergravity
backgrounds.  We discuss the Minkowski vacua in all three theories as
well as the elementary brane backgrounds.  As in the analysis of
Kaluza--Klein reductions in \cite{FigSimBranes}, the asymptotic
flatness of the brane backgrounds allows us to rephrase questions
about the symmetries of these backgrounds in terms of the symmetries
of the asymptotic Minkowski vacuum.  In particular their Killing
superalgebras are subsuperalgebras of the relevant Poincaré
superalgebra and the computation of deformations will borrow much from
the case of the Poincaré superalgebras.

We will not explain the methodology in this paper.  It is explained in
\cite[Section~2]{JMFSuperDeform}, which the reader should consult for
the details.  In a nutshell, the tangent space to the moduli space of
deformations of a Lie superalgebra $\fk$ is given by the cohomology
group $H^2(\fk;\fk)$, which we calculate for the Killing superalgebras
of these supergravity backgrounds by using the superalgebra version of
the factorisation theorem of Hochschild and Serre.  If $H^2(\fk;\fk) =
0$ we say that $\fk$ is \emph{rigid}.  Otherwise, every line in
$H^2(\fk;\fk)$ defines an infinitesimal deformation of $\fk$ and one
can investigate whether it integrates to a one-parameter deformation.
This requires the vanishing of a potentially infinite number of
obstructions in $H^3(\fk;\fk)$, but in practice we will not have to go
beyond second order in any of the deformations found here.

The complexity of the calculations increases as we move from type I to
type IIB and then to type IIA supergravities, and we have decided to
organise the paper in increasing complexity.  Within each theory,
however, we have ordered the sections in such a way that we first
treat the Minkowski and Freund--Rubin vacua and then the brane-like
backgrounds in increasing brane dimension.  We now give a summary of
the results.

In Section~\ref{sec:TypeI} we discuss type I backgrounds.  In
Section~\ref{sec:PoincareI} we prove the rigidity of the Poincaré
superalgebra and in Section~\ref{sec:ID5} we prove that of the
D5-brane superalgebra, whereas in Section~\ref{sec:ID1} we exhibit a
one-parameter deformation of the D1-brane superalgebra.  In
Section~\ref{sec:TypeIIB} we discuss type IIB backgrounds.  The
rigidity of the Poincaré superalgebra is demonstrated in
Section~\ref{sec:PoincareIIB}, whereas in Section~\ref{sec:MaxSusyIIB}
we discuss other maximally supersymmetric backgrounds.  We sketch a
proof that the superalgebra of the Freund--Rubin background is rigid,
whereas the existence of the plane-wave limit shows that the
superalgebra of the maximally supersymmetric wave admits at least one
deformation.  In Section~\ref{sec:IIBD1} we exhibit a one-parameter
deformation of the D1-brane superalgebra, whereas in
Sections~\ref{sec:D3} and \ref{sec:IIBD5}, we prove the rigidity of
the D3- and D5-brane superalgebras, respectively.  The rigidity of the
D3-brane superalgebra may come as a surprise in view of the
deformation of the four-dimensional Poincaré superalgebra.  For
completeness and because there seems to be some confusion in the
literature on this topic, we work out this deformation in
Section~\ref{sec:4d}.  In Section~\ref{sec:IIBF1} and \ref{sec:D7},
respectively, we exhibit one-parameter deformations of the
superalgebras of the D7-brane and of the fundamental string, whereas
in Section~\ref{sec:IIBNS5} we prove the rigidity of the superalgebra
of the NS5-brane.  In Section~\ref{sec:TypeIIA} we discuss type IIA
supergravity backgrounds.  The rigidity of the Poincaré superalgebra
is shown in Section~\ref{sec:PoincareIIA}, whereas the rigidity of the
superalgebras of the D0-, D4- and NS5-branes is shown in
Sections~\ref{sec:D0}, \ref{sec:D4} and \ref{sec:IIANS5},
respectively.  We exhibit deformations of the fundamental string, D2-
and D6-brane superalgebras in Sections~\ref{sec:IIAF1}, \ref{sec:D2}
and \ref{sec:D6}, respectively.  The latter two deformations have
their origin in the deformations of the superalgebras of the
delocalised M2-brane and the Kaluza--Klein monopole in
eleven-dimensional supergravity.  The latter deformation was found in
\cite{JMFSuperDeform}, whereas the former is described in
Section~\ref{sec:DM2}.  Finally in Section~\ref{sec:conclusions} we
summarise our results and speculate on the geometric origin of these
deformations.  The paper ends with Appendix~\ref{sec:spin} which lists
our spinor conventions and records some useful formulae.

\section{Type I backgrounds}
\label{sec:TypeI}

In this section we study the Lie superalgebra deformations of the
Killing superalgebra \cite{EHJGMHom} of some type I supergravity
backgrounds: Minkowski space, which is the unique maximally
supersymmetric background, and the half-BPS D1- and D5-brane
backgrounds.

\subsection{Rigidity of the Poincaré superalgebra}
\label{sec:PoincareI}

The Killing superalgebra of the Minkowski vacuum of type I
supergravity is the type I Poincaré superalgebra.  Let $V$ denote
a ten-dimensional lorentzian vector space and let $\fso(V)$ denote the
corresponding Lorentz Lie algebra.  The Poincaré Lie algebra is
$\fso(V) \oplus V$.  The type I spinors are chiral, and we take them
to have positive chirality without loss of generality.  Let $\Delta_+$
denote their representation space.  As a vector superspace, the
type I Poincaré superalgebra is $\fk = \fk_0 \oplus \fk_1$ with
$\fk_0 \cong \fso(V) \oplus V$ and $\fk_1 \cong \Delta_+$.  Let
$\be_\mu$ denote an orthonormal frame for $V$, $\be_\mu \wedge
\be_\nu$, for $\mu < \nu$ a basis for $\Lambda^2V$ and let
$\beps_\alpha$ denote a basis for $\Delta_+$.  The corresponding basis
for $\fk$ will be denoted $P_\mu$, $L_{\mu\nu}$ and $Q_\alpha$.  The
supertranslation ideal $I < \fk$ is spanned by $P_\mu$ and $Q_\alpha$,
whereas the semisimple factor $\fs$ is the span of the $L_{\mu\nu}$.
The Lie brackets are those of the Lorentz subalgebra and in addition
\begin{equation}
  \label{eq:IPoincare}
  \begin{aligned}[m]
    [L_{\mu\nu},Q_\alpha] &= \half \Gamma_{\mu\nu} \cdot Q_\alpha\\
    [L_{\mu\nu},P_\rho] &=  \eta_{\nu\rho} P_\mu - \eta_{\mu\rho} P_\nu\\
    [Q_\alpha, Q_\beta] &= \Gamma^\mu_{\alpha\beta} P_\mu~,
  \end{aligned}
\end{equation}
with $\eta_{\mu\nu}$ the Minkowski metric relative to this orthonormal
frame, and where
\begin{equation}
  \Gamma_{\mu\nu} \cdot Q_\alpha = Q_\beta
  (\Gamma_{\mu\nu})^\beta{}_\alpha~,
\end{equation}
and similarly for the action of any other element in the Clifford
algebra $\Cl(V)$, and
\begin{equation}
  \Gamma^\mu_{\alpha\beta} := \left<\beps_\alpha, \Gamma^\mu  \beps_\beta\right>~,
\end{equation}
where $\left<-,-\right>$ is the $\fs$-invariant symplectic structure
on $\Delta$.

We are interested in the cohomology group $H^2(\fk;\fk)$ which can be
computed from the complex $C^\bullet := C^\bullet(I;\fk)^{\fs}$ of
$\fs$-equivariant linear maps $\Lambda^\bullet I \to \fk$.  We write
these maps tensorially as invariant elements in $\Lambda^\bullet I^*
\otimes \fk$.  It should be pointed out that this way of writing them
incurs in some signs.  Indeed, whereas the natural isomorphism
$\Hom(\Lambda^\bullet I,\fk) \cong \fk \otimes \Lambda^\bullet I^*$
carries no sign, the isomorphism $\fk \otimes \Lambda^\bullet I^*
\cong \Lambda^\bullet I^* \otimes \fk$ does carry signs whenever we
are interchanging odd objects.  Let $P^\mu$ and $Q^\alpha$ denote the
canonical dual basis for $I^*$.  The differential $d$ of the complex
$C^\bullet$ is defined uniquely by the following action on $I^*$ and
on $\fk$ as an $I$-module:
\begin{equation}
  \begin{aligned}[m]
    d P^\mu &= \half \Gamma^\mu_{\alpha\beta} Q^\alpha \wedge Q^\beta\\
    d Q^\alpha &= 0\\
    d P_\mu &= 0\\
    d Q_\alpha &= - \Gamma^\mu_{\alpha\beta} Q^\beta \otimes  P_\mu\\
    d L_{\mu\nu} &= \eta_{\mu\rho} P^\rho \otimes P_\nu - \eta_{\nu\rho} P^\rho \otimes P_\mu + \half
    Q^\alpha \otimes \Gamma_{\mu\nu} \cdot Q_\alpha~.
  \end{aligned}
\end{equation}

As there are no Lorentz scalars in $\fk$, $C^0 = 0$.  There are also
no $1$-coboundaries.  The space $C^1$ of $1$-cochains is spanned by
the cochains corresponding to the identity maps $V \to V$ and
$\Delta_+ \to \Delta_+$; that is, $P^\mu \otimes P_\mu$ and $Q^\alpha
\otimes Q_\alpha$.  Computing the differential $d: C^1 \to C^2$, we
find
\begin{equation}
  \begin{aligned}[m]
    d \left(P^\mu \otimes P_\mu\right) &= \half
    \Gamma^\mu_{\alpha\beta} Q^\alpha \wedge Q^\beta \otimes  P_\mu\\
    d \left(Q^\alpha \otimes Q_\alpha\right) &=
    \Gamma^\mu_{\alpha\beta} Q^\alpha \wedge Q^\beta \otimes P_\mu~,
  \end{aligned}
\end{equation}
whence we see that there is one cocycle $2 P^\mu \otimes P_\mu -
Q^\alpha \otimes Q_\alpha$.  We conclude that $H^1(\fk;\fk) \cong
\RR$, corresponding to the outer derivation which gives $Q_\alpha$ weight
$1$, $P_\mu$ weight $2$, and $L_{\mu\nu}$ weight $0$.  We also see
that there is a one-dimensional space of $2$-coboundaries, spanned by
$\Gamma^\mu_{\alpha\beta} Q^\alpha \wedge Q^\beta \otimes P_\mu$.

The space of $2$-cochains consists of $\fs$-equivariant maps
$\Lambda^2 I \to \fk$.  As there are no such maps $V \otimes \Delta_+
\to \Delta_+$ and $\Delta_+ \otimes \Delta_+ \to \Lambda^2 V$, we find
only the following $2$-cochains: $P^\mu\wedge P^\nu \otimes
L_{\mu\nu}$ and $\Gamma^\mu_{\alpha\beta} Q^\alpha \wedge Q^\beta
\otimes P_\mu$, corresponding to the natural isomorphism $\Lambda^2V
\to \fso(V)$ and the projection $S^2\Delta_+ \to V$.  A simple
calculation shows that
\begin{equation}
  d \left( P^\mu\wedge P^\nu \otimes L_{\mu\nu} \right) =
  \Gamma^\mu_{\alpha\beta} P^\nu \wedge Q^\alpha \wedge Q^\beta
  \otimes L_{\mu\nu} + \half P^\mu\wedge P^\nu \wedge Q^\alpha \otimes
  \Gamma_{\mu\nu} \cdot Q_\alpha \neq 0~,
\end{equation}
whence the only cocycle is also a coboundary and hence $H^2(\fk;\fk) =
0$, showing the rigidity of the type I Poincaré superalgebra.

\subsection{A deformation of the D1-brane superalgebra}
\label{sec:ID1}

To describe the type I D1-brane superalgebra, we split the
ten-dimensional space as $V = W \oplus W^\perp$, where $W$ is
lorentzian and two-dimensional.  The D1 superalgebra is $\fk = \fk_0
\oplus \fk_1$ with $\fk_0 = \fso(W) \oplus W \oplus \fso(W^\perp)$ and
$\fk_1 \cong \Delta_{\D1} = \left\{\varepsilon \in \Delta_+ \middle |
  \bnu_W  \varepsilon = \varepsilon \right\}$, where the volume
element $\bnu_W$ of $W$ is skewsymmetric relative to the spinor inner
product and satisfies $\bnu_W^2 = + \1$.  Let $L_{\mu\nu} =
\epsilon_{\mu\nu} L$, $P_\mu$, $L_{ab}$ and $Q_\alpha$ be a basis for
$\fk$.  The nonzero Lie brackets are those of $\fso(W^\perp)$ and in
addition
\begin{equation}
  \label{eq:ID1KSA}
  \begin{aligned}[m]
    [L,Q_\alpha] &= - \half Q_\alpha\\
    [L_{ab},Q_\alpha] &= \half \Gamma_{ab} \cdot Q_\alpha\\
    [L,P_\mu] &=  \epsilon_\mu{}^\nu P_\nu\\
    [Q_\alpha, Q_\beta] &= \Gamma^\mu_{\alpha\beta} P_\mu~.
  \end{aligned}
\end{equation}

Since $\fso(W)$ is abelian, it is not part of the semisimple factor
and it must be included in the ideal $I$, which is now spanned by
$P_\mu$, $Q_\alpha$ and $L$.  The semisimple subalgebra $\fs$ is now
$\fso(W^\perp)$.  However $\fr := \fso(W) \oplus \fso(W^\perp)$ is
reductive and a slight refinement of the Hochschild--Serre
factorisation theorem allows us to work with cochains which are
invariant under $\fr$.  Indeed, the cohomology group $H^2(\fk;\fk)$
can be computed from the complex $C^\bullet := C^\bullet(I;\fk)^{\fr}$
of $\fr$-equivariant linear maps $\Lambda^\bullet I \to \fk$.  Letting
$L^*,P^\mu,Q^\alpha$ denote the canonical dual basis for $I^*$, the
differential $d$ of this complex is defined uniquely by the following
relations
\begin{equation}
  \begin{aligned}[m]
    d P^\mu &= \half \Gamma^\mu_{\alpha\beta} Q^\alpha \wedge Q^\beta
    + \epsilon^\mu{}_\nu L^* \wedge P^\nu \\
    d Q^\alpha &= \half L^* \wedge Q^\alpha\\
    d L^* &= 0\\
    d P_\mu &= L^* \otimes \epsilon_\mu{}^\nu P_\nu\\
    d Q_\alpha &= -\half L^* \otimes Q_\alpha - \Gamma^\mu_{\alpha\beta} Q^\beta \otimes  P_\mu\\
    d L &= - P^\mu \otimes \epsilon_\mu{}^\nu P_\nu - \half Q^\alpha
    \otimes Q_\alpha\\
    d L_{ab} &= \half Q^\alpha \otimes \Gamma_{ab} \cdot Q_\alpha~.
  \end{aligned}
\end{equation}

In this case, $C^0 = \fk^\fr$ is spanned by $L$, but since $dL \neq
0$, $H^0(\fk;\fk) = 0$ and $\dim B^1 = 1$.  The space of $1$-cochains
is $4$-dimensional, with basis $L^* \otimes L$, $P^\mu \otimes P_\mu$,
$P^\mu \otimes \epsilon_\mu{}^\nu P_\nu$ and $Q^\alpha \otimes
Q_\alpha$.  Finally, the space of $2$-cochains is $5$-dimensional,
spanned by $P^\mu\wedge P^\nu \otimes \epsilon_{\mu\nu} L$, $L^*
\wedge P^\mu \otimes P_\mu$, $L^* \otimes P^\mu \otimes
\epsilon_\mu{}^\nu P_\nu$, $L^* \wedge Q^\alpha \otimes Q_\alpha$ and
$Q^\alpha \wedge Q^\beta \otimes \Gamma^\mu_{\alpha\beta} P_\mu$.
Computing the differentials $d:C^1 \to C^2$ and $d: C^2 \to C^3$, we
find that $H^1(\fk;\fk) \cong \RR$, with representative cocycle
$\varphi:= 2P^\mu \otimes P_\mu - Q^\alpha \otimes Q_\alpha$, and
$H^2(\fk;\fk) \cong \RR$, with representative cocycle $L^* \wedge
\varphi$.  This infinitesimal deformation integrates to a
one-parameter family of Lie superalgebras with brackets
\begin{equation}
  \label{eq:ID1deformed}
  \begin{aligned}[m]
    [L,Q_\alpha] &= (t - \half) Q_\alpha\\
    [L,P_\mu] &=  2 t P_\mu + \epsilon_\mu{}^\nu P_\nu\\
    [Q_\alpha, Q_\beta] &= \Gamma^\mu_{\alpha\beta} P_\mu
  \end{aligned}
\end{equation}
in addition to those involving $\fso(W^\perp)$, which remain
undeformed.  The reader may be forgiven for suspecting a discrepancy
from the sign of the $t$-dependent terms in the brackets above and the
relative sign in the cocycle $\varphi$.  As explained above, this is
due to the signs in the isomorphism $W \otimes V^* \cong V^* \otimes
W$ whenever $V$ and $W$ are both odd subspaces.  The signs can be read
off from the formulae in \cite[Section~2]{JMFSuperDeform}.

\subsection{Rigidity of the D5-brane superalgebra}
\label{sec:ID5}

The type I D5-brane superalgebra is the subsuperalgebra of the Poincaré
superalgebra defined as follows.  We first split the ten-dimensional
lorentzian vector space $V = W \oplus W^\perp$, where $W$ is a
six-dimensional lorentzian subspace.  Then the D5-brane superalgebra
$\fk = \fk_0 \oplus \fk_1$, where $\fk_0 = \fso(W) \oplus W \oplus
\fso(W^\perp)$ and $\fk_1 \cong \Delta_{\D5} = \left\{\varepsilon \in
\Delta_+ \middle | \bnu_W  \varepsilon = \varepsilon \right\}$,
where $\bnu_W$ is the Clifford algebra element which represents the
volume form of $W$.  It is skewsymmetric relative to the invariant
symplectic form on spinors and satisfies $\bnu_W^2 = + \1$.  Let
$\varepsilon_\alpha$ be a basis for $\Delta_{\D5}$ and $\be_\mu$ and
$\be_a$ bases for $W$ and $W^\perp$, respectively.  Let $Q_\alpha$ and
$P_\mu$ denote the corresponding basis for the ideal $I < \fk$.  The
semisimple subalgebra $\fs$ is spanned by $L_{\mu\nu}$ and $L_{ab}$.
Relative to this basis, the Lie brackets of $\fk$ are given by those
of $\fso(W) \oplus \fso(W^\perp)$ and
\begin{equation}
  \label{eq:ID5KSA}
  \begin{aligned}[m]
    [L_{\mu\nu},Q_\alpha] &= \half \Gamma_{\mu\nu} \cdot Q_\alpha\\
    [L_{ab},Q_\alpha] &= \half \Gamma_{ab} \cdot Q_\alpha\\
    [L_{\mu\nu},P_\rho] &=  \eta_{\nu\rho} P_\mu - \eta_{\mu\rho} P_\nu\\
    [Q_\alpha, Q_\beta] &= \Gamma^\mu_{\alpha\beta} P_\mu~,
  \end{aligned}
\end{equation}
where again
\begin{equation}
  \Gamma_{\mu\nu} \cdot Q_\alpha = Q_\beta (\Gamma_{\mu\nu})^\beta{}_\alpha
\end{equation}
and
\begin{equation}
  \Gamma^\mu_{\alpha\beta} := \left<\beps_\alpha, \Gamma^\mu  \beps_\beta\right>~.
\end{equation}

Let $P^\mu$ and $Q^\alpha$ denote the canonical dual
basis for $I^*$.  The differential $d$ of the complex $C^\bullet = 
C^\bullet(I;\fk)^{\fs}$ is defined uniquely by
\begin{equation}
  \begin{aligned}[m]
    d P^\mu &= \half \Gamma^\mu_{\alpha\beta} Q^\alpha \wedge Q^\beta\\
    d Q^\alpha &= 0\\
    d P_\mu &= 0\\
    d Q_\alpha &= - \Gamma^\mu_{\alpha\beta} Q^\beta \otimes  P_\mu\\
    d L_{\mu\nu} &= \eta_{\mu\rho} P^\rho \otimes P_\nu - \eta_{\nu\rho} P^\rho \otimes P_\mu + \half
    Q^\alpha \otimes \Gamma_{\mu\nu} \cdot Q_\alpha\\
    d L_{ab} &= \half Q^\alpha \otimes \Gamma_{ab} \cdot Q_\alpha~.
  \end{aligned}
\end{equation}

Again we see that $C^0 = \fk^{\fs} = 0$ and that $C^1$ is spanned by
$P^\mu \otimes P_\mu$ and $Q^\alpha \otimes Q_\alpha$.  Similarly,
$C^2$ is spanned by $P^\mu \wedge P^\nu \otimes L_{\mu\nu}$ and
$Q^\alpha \wedge Q^\beta \otimes \Gamma^\mu_{\alpha\beta} P_\mu$.
It is easy to compute the differentials $d:C^1 \to C^2$ and $d:C^2 \to
C^3$ and we see that $H^1(\fk;\fk) \cong \RR$, with representative
cocycle $2P^\mu \otimes P_\mu - Q^\alpha\otimes Q_\alpha$, and that
$H^2(\fk;\fk) = 0$, whence the type I D5-brane superalgebra is rigid.

\section{Type IIB backgrounds}
\label{sec:TypeIIB}

In this section we explore the Lie superalgebra deformations of the
Killing superalgebras of certain type IIB backgrounds.  We start with
the Minkowski vacuum, treat briefly the other maximally supersymmetric
backgrounds and then go on to the elementary brane backgrounds.

\subsection{Rigidity of the Poincaré superalgebra}
\label{sec:PoincareIIB}

The Killing superalgebra of the Minkowski vacuum is the type IIB
Poincaré superalgebra $\fk = \fk_0 \oplus \fk_1$, with $\fk_0 =
\fso(V) \oplus V$ the Poincaré algebra and $\fk_1$ isomorphic to two
copies of the positive chirality spinor representation $\Delta_+$ of
$\fso(V)$.  We will denote these two copies by $\Delta_+^I$, where
$I=1,2$.  Let $\beps_\alpha$ be a basis for $\Delta_+$ and let
$\beps^I_\alpha$ and $Q_\alpha^I$ denote the corresponding bases for
$\Delta_+^I$ and $\fk_1$, respectively.  Then $\fk$ is spanned by
$P_\mu$, $L_{\mu\nu}$, and $Q_\alpha^I$ subject to the following
brackets, in addition to the ones of the Poincaré subalgebra,
\begin{equation}
  \label{eq:IIBPoincare}
  \begin{aligned}[m]
    [L_{\mu\nu},Q^I_\alpha] &= \half \Gamma_{\mu\nu} \cdot Q^I_\alpha\\
    [Q^I_\alpha, Q^J_\beta] &= \delta^{IJ}\Gamma^\mu_{\alpha\beta} P_\mu~,
  \end{aligned}
\end{equation}
where
\begin{equation}
  \Gamma_{\mu\nu} \cdot Q^I_\alpha = Q^I_\beta
  (\Gamma_{\mu\nu})^\beta{}_\alpha~,
\end{equation}
and
\begin{equation}
  \Gamma^\mu_{\alpha\beta} := \left<\beps_\alpha, \Gamma^\mu 
    \beps_\beta\right>~.
\end{equation}
In other words, $\Cl(V)$ and the spinor inner product act
independently in each of the two copies of the spinor representation.

The supertranslation ideal $I < \fk$ is now spanned by $Q_\alpha^I$
and $P_\mu$ whereas the semisimple factor $\fs = \fso(V)$.  Let
$Q_I^\alpha$ and $P^\mu$ denote the canonical dual basis for $I^*$.

Let $C^\bullet = C^\bullet(I;\fk)^{\fs}$ denote the complex of
$\fs$-invariant maps $\Lambda^\bullet I \to \fk$, with differential
$d$ defined by the following relations
\begin{equation}
  \begin{aligned}[m]
    d P^\mu &= \half \Gamma^\mu_{\alpha\beta} Q^\alpha \wedge Q^\beta\\
    d Q_I^\alpha &= 0\\
    d P_\mu &= 0\\
    d Q^I_\alpha &= - \delta^{IJ} \Gamma^\mu_{\alpha\beta} Q_J^\beta \otimes  P_\mu\\
    d L_{\mu\nu} &= \eta_{\mu\rho} P^\rho \otimes P_\nu - \eta_{\nu\rho} P^\rho \otimes P_\mu + \half
    Q_I^\alpha \otimes \Gamma_{\mu\nu} \cdot Q^I_\alpha~.
  \end{aligned}
\end{equation}
As there are no Lorentz scalars in $\fk$, $C^0 = \fk^{\fs} = 0$.
The space $C^1$ of $1$-cochains is $5$-dimensional, spanned by
the cochains corresponding to the identity maps $V \to V$ and
$\Delta_+^I \to \Delta_+^J$, namely $P^\mu \otimes P_\mu$ and
$Q^\alpha_I \otimes Q^J_\alpha$.  The space $C^2$ is $4$-dimensional,
spanned by the natural isomorphism $\Lambda^2V \cong \fso(V)$ and the
projections $\Delta_+^I \otimes \Delta_+^J \to V$ which are symmetric
in $I\leftrightarrow J$.  Evaluating the differential $d: C^1 \to C^2$
we find
\begin{equation}
  \begin{aligned}[m]
    d \left(P^\mu \otimes P_\mu\right) &= \half
    \delta^{IJ} \Gamma^\mu_{\alpha\beta} Q_I^\alpha \wedge Q_J^\beta
    \otimes  P_\mu\\
    d \left(Q_I^\alpha \otimes Q^J_\alpha\right) &=
    \delta^{JK} \Gamma^\mu_{\alpha\beta} Q_I^\alpha \wedge Q_K^\beta \otimes P_\mu~,
  \end{aligned}
\end{equation}
from where we see that $H^1(\fk;\fk) \cong \RR^2$, with representative
cocycles $2 P^\mu \otimes P_\mu - Q^\alpha_I \otimes Q^I_\alpha$ and
$\epsilon_J{}^I Q^\alpha_I \otimes Q^J_\alpha$.  This means that $\dim
B^2 = 3$ and since
\begin{equation}
  d\left(P^\mu \wedge P^\nu \otimes L_{\mu\nu}\right) = \delta^{IJ}
  \Gamma^\mu_{\alpha\beta} Q_I^\alpha \wedge Q_J^\beta \wedge P^\nu
  \otimes L_{\mu\nu} + \half P^\mu \wedge P^\nu \wedge Q^\alpha_I
  \otimes \Gamma_{\mu\nu} \cdot Q_\alpha^I  \neq 0~,
\end{equation}
we see that $H^2(\fk;\fk) = 0$ and the IIB Poincaré superalgebra is
rigid.

\subsection{Other maximally supersymmetric backgrounds}
\label{sec:MaxSusyIIB}

As shown in \cite{FOPMax}, there are only two other maximally
supersymmetric IIB backgrounds: the Freund--Rubin background
\cite{SchwarzIIB} with geometry $\AdS_5 \times S^5$, and the maximally
supersymmetric wave \cite{NewIIB}.  The Killing superalgebra of the
Freund--Rubin background is the simple Lie superalgebra $\fsu(2,2|4)$,
whereas that of the maximally supersymmetric wave is the contraction
\cite{ShortLimits,Limits,HatKamiSaka} induced by the plane-wave limit
\cite{PenrosePlaneWave,GuevenPlaneWave}.  This observation implies
that the Killing superalgebra of the maximally supersymmetric wave is
not rigid, and it admits at least a one-parameter family of
deformations, isomorphic to $\fsu(2,2|4)$ for nonzero values of the
parameter.  We will not compute the space of deformations in this
paper, but as in the similar situation in eleven dimensions
\cite{JMFSuperDeform}, we would be surprised if there were any other
deformations.

As for $\fsu(2,2|4)$ itself, the fact that it is simple does not
immediately imply that it is rigid.  A closer look at the rigidity
results for simple Lie superalgebras \cite{SSSLjahovski} shows that
the crucial condition used in the proof is the nondegeneracy of the
Killing form.  Whereas Cartan's criterion guarantees that this is the
case for semisimple Lie algebras, this is not the case for
superalgebras.  Indeed, in Kac's list \cite{KacSuperSketch} there are
simple Lie superalgebras with degenerate (or even zero) Killing form
and indeed, the Lie superalgebra of type $D(2,1)$ has zero Killing
form and admits a one-parameter deformation $D(2,1;\alpha)$ which
remains simple for all values of $\alpha$.  Curiously, as shown in
\cite{GMT1}, the Killing superalgebra of the near-horizon geometry of
a $\tfrac18$-BPS configuration of rotating intersecting branes in
eleven-dimensional supergravity is isomorphic to two copies of
$D(2,1;\alpha)$---the parameter $\alpha$ having a geometric
interpretation as the ratio of the radii of the two $3$-spheres in the
near-horizon geometry $\AdS_3 \times S^3 \times S^3 \times \RR^2$.

The Lie superalgebra $\fsu(2,2|4)$ too has zero Killing form, hence
the result of \cite{SSSLjahovski} does not apply, and moreover since
the algebra is simple, there is no Hochschild--Serre factorisation.
However the cohomology $H^2(\fk;\fk)$ may be calculated from the
subcomplex of cochains which are invariant under the semisimple even
subalgebra $\fk_0 = \fsu(2,2) \oplus \fsu(4)$.  The space of
$2$-cochains breaks up into three, which are the $\fk_0$-invariant
subspaces of $\Lambda^2 \fk_0^* \otimes \fk_0$, $\fk_0^* \otimes
\fk_1^* \otimes \fk^1$ and $S^2\fk_1^* \otimes \fk_0$.  As an
$\fk_0$-module, $\fk_0 = (\Lambda^2V^{(2,4)}\otimes \RR) \oplus (\RR
\otimes \Lambda^2 V^{(6)})$, where $V^{(2,4)}$ and $V^{(6)}$ are the
vector representations of $\fso(2,4) \cong \fsu(2,2)$ and $\fso(6)
\cong \fsu(4)$, respectively.  Similarly, $\fk_1 = [\![ \Delta^{(2,4)}
\otimes \Delta^{(6)}]\!]$, where $\Delta^{(2,4)}$ and $\Delta^{(6)}$
are the positive-chirality spinor representations of $\fso(2,4)$ and
$\fso(6)$, respectively, which are complex four-dimensional, and where
if $W$ is a complex vector space, $[\![ W ]\!]$ is a real vector space
defined by $[\![ W ]\!]\otimes_\RR \CC = W \oplus \bar W$.  In other
words, it is the vector space spanned by the real and imaginary parts
of the vectors in $W$, whence $\dim_\RR [\![ W ]\!] = 2 \dim_\CC W$.
In this case, $\Delta^{(2,4)}\otimes \Delta^{(6)}$ is complex and
$16$-dimensional, whence $\fk_1$ is real and $32$-dimensional, as
expected.  Since $\fk_0$ is semisimple, it is rigid as a Lie algebra,
we can assume that its Lie brackets remain undeformed, hence we can
assume that a cocycle defining an infinitesimal deformation of
$\fsu(2,2|4)$ has no components in $\Lambda^2 \fk_0^* \otimes \fk_0$.
By the same token, the rigidity of $\fk_1$ as an $\fk_0$-module says
that the putative cocycle cannot have components in $\fk_0^* \otimes
\fk_1^* \otimes \fk^1$, whence the cocycle, if it exists, must belong
to the $\fk_0$-invariant subspace of $S^2\fk_1^* \otimes \fk_0$.  A
simple roots-and-weights calculation shows that this space is
two-dimensional made out of the natural maps
\begin{gather}
  \Delta^{(2,4)}\otimes \bar\Delta^{(2,4)} \to \RR \qquad\text{and}\qquad
  \Delta^{(6)}\otimes \bar\Delta^{(6)} \to \Lambda^2 V^{(6)}\\
  \Delta^{(6)}\otimes \bar\Delta^{(6)} \to \RR \qquad\text{and}\qquad
  \Delta^{(2,4)}\otimes \bar\Delta^{(2,4)} \to \Lambda^2 V^{(2,4)}
\end{gather}
which means that the $[\fk_1,\fk_1]$ bracket has two parameters, which
we can choose to take the value $1$ in the undeformed superalgebra
$\fsu(2,2|4)$.  The $(\fk_1,\fk_1,\fk_1)$ Jacobi identity fixes the
ratio of these two parameters to be $1$ and we can further set them to
be equal to $1$ by rescaling the odd generators, hence proving the
rigidity of the superalgebra.

\subsection{A deformation of the D1-brane superalgebra}
\label{sec:IIBD1}

The Killing superalgebra of the type IIB D1-brane is the
subsuperalgebra $\fk$ of the IIB Poincaré superalgebra with $\fk_0 =
\fso(W) \oplus W \oplus \fso(W^\perp)$, where $V = W \oplus W^\perp$
is the decomposition of the $10$-dimensional lorentzian vector space
into a $2$-dimensional lorentzian subspace $W$, corresponding to the
brane worldvolume and its $8$-dimensional perpendicular complement.
The odd subspace $\fk_1$ is isomorphic to the graph $\Delta_{\D1}
\subset \Delta_+ \oplus \Delta_+$ of the endomorphism $\bnu_W:
\Delta_+ \to \Delta_+$ corresponding to the volume form of $W$.  As in
the type I D1-brane, the extension of $\bnu_W$ to the Clifford module
is skewsymmetric relative to the spinor inner product and obeys
$\bnu_W^2 = + \1$.

Let $\be_\mu$ and $\be_a$ span $W$ and $W^\perp$, respectively and let
$\beps_\alpha$ span $\Delta_+$.  Let $\bpsi_\alpha =
\tfrac1{\sqrt{2}} \begin{pmatrix}\beps_\alpha\\\bnu_W 
  \beps_\alpha\end{pmatrix}$ be a basis for $\Delta_{\D1}$.  The
corresponding basis of $\fk$ is given by $P_\mu$, $L_{\mu\nu} =
\epsilon_{\mu\nu}L$, $L_{ab}$ and $Q_\alpha$.  The Lie brackets are
inherited from those in equation \eqref{eq:IIBPoincare} and are given
explicitly, in addition to those involving $L_{ab}$, by
\begin{equation}
  \label{eq:IIBD1KSA}
  \begin{aligned}[m]
    [L,Q_\alpha] &= - \half \bnu_W \cdot Q_\alpha\\
    [L,P_\mu] &=  \epsilon_\mu{}^\nu P_\nu\\
    [Q_\alpha, Q_\beta] &= \Gamma^\mu_{\alpha\beta} P_\mu~,
  \end{aligned}
\end{equation}
where
\begin{equation}
  \Gamma^\mu_{\alpha\beta} := \langle\bpsi_\alpha, \Gamma^\mu 
    \bpsi_\beta\rangle =  \left<\beps_\alpha, \Gamma^\mu 
    \beps_\beta\right>~.
\end{equation}

As in the case of the type I D1-brane calculation, the ideal $I < \fk$
includes the generator $L$, but we may work with cochains which are
invariant under the reductive subalgebra $\fr := \fso(W) \oplus
\fso(W^\perp)$.  Letting $L^*$, $P^\mu$ and $Q^\alpha$ be a basis for
$I^*$, the differential in the complex $C^\bullet =
C^\bullet(I,\fk)^{\fr}$ is determined by the following relations
\begin{equation}
  \begin{aligned}[m]
    d P^\mu &= \half \Gamma^\mu_{\alpha\beta} Q^\alpha \wedge Q^\beta
    + \epsilon^\mu{}_\nu L^* \wedge P^\nu \\
    d Q^\alpha &= \half L^* \wedge \left(\bnu_W\right)^\alpha{}_\beta Q^\beta\\
    d L^* &= 0\\
    d P_\mu &= L^* \otimes \epsilon_\mu{}^\nu P_\nu\\
    d Q_\alpha &= -\half L^* \otimes \bnu_W \cdot Q_\alpha - \Gamma^\mu_{\alpha\beta} Q^\beta \otimes  P_\mu\\
    d L &= - P^\mu \otimes \epsilon_\mu{}^\nu P_\nu - \half Q^\alpha
    \otimes \bnu_W \cdot Q_\alpha\\
    d L_{ab} &= \half Q^\alpha \otimes \Gamma_{ab} \cdot Q_\alpha~.
  \end{aligned}
\end{equation}

The $0$-cochains $C^0 = \fk^{\fr}$ are spanned by $L$, but $dL \neq
0$, hence $H^0(\fk;\fk) = 0$ and $\dim B^1 = 1$.  The space of
$1$-cochains is $5$-dimensional, spanned by $P^\mu \otimes P_\mu$,
$P^\mu \otimes \epsilon_\mu{}^\nu P_\nu$, $L^*\otimes L$, $Q^\alpha
\otimes Q_\alpha$ and $Q^\alpha \otimes \bnu_W \cdot Q_\alpha$.
Another way to understand this is to notice that the ideal $I$ is
graded by the action of $2L$ with $L$ having degree $0$, $Q_\alpha$
having pieces of degrees $\pm 1$ and $P_\mu$ having pieces of degrees
$\pm 2$, corresponding to a Witt basis for $W$.  The $5$-dimensional
space of cochains can be thought of as spanned by the cochains
corresponding to the identity maps of each of the five graded
subspaces.  The space of $2$-cochains is $7$-dimensional, spanned by
$P^\mu\wedge P^\nu\otimes \epsilon_{\mu\nu} L$, $L^* \wedge P^\mu
\otimes P_\mu$, $L^* \wedge P^\mu \otimes \epsilon_\mu{}^\nu P_\nu$,
$L^* \wedge Q^\alpha \otimes Q_\alpha$, $L^* \wedge Q^\alpha
\otimes \bnu_W \cdot Q_\alpha$, $Q^\alpha \wedge Q^\beta \otimes
\Gamma^\mu_{\alpha\beta} P_\mu$ and $Q^\alpha \wedge Q^\beta \otimes
\left(\Gamma^\mu\bnu_W\right)_{\alpha\beta} P_\mu$.

Computing the differential $d: C^1 \to C^2$, we find that
$H^1(\fk;\fk) \cong \RR$, with representative cocycle $\varphi:= 2
P^\mu \otimes P_\mu - Q^\alpha \otimes Q_\alpha$.  Similarly,
computing $d:C^2 \to C^3$ we find that $H^2(\fk;\fk)\cong \RR$, with
representative cocycle $L^* \wedge \varphi$.  This infinitesimal
deformation integrates to a one-parameter family of Lie superalgebras
with brackets
\begin{equation}
  \label{eq:IIBD1deformed}
  \begin{aligned}[m]
    [L,Q_\alpha] &= t Q_\alpha - \half \bnu_W \cdot Q_\alpha\\
    [L,P_\mu] &=  2 t P_\mu + \epsilon_\mu{}^\nu P_\nu\\
    [Q_\alpha, Q_\beta] &= \Gamma^\mu_{\alpha\beta} P_\mu
  \end{aligned}
\end{equation}
in addition to those involving $\fso(W^\perp)$, which remain
undeformed.  This deformation consists of changing the $L$-weight of
the generators in the Lie superalgebra in such a way that the $QQ$
bracket remains invariant.  This deformation is familiar from the
twisting construction of two-dimensional topological conformal field
theories.

\subsection{A deformation of the fundamental string superalgebra}
\label{sec:IIBF1}

The Killing superalgebra of the type IIB fundamental string is the
subsuperalgebra $\fk = \fk_0 \oplus \fk_1$ of the IIB Poincaré
superalgebra where $\fk_0 = \fso(W) \oplus W \oplus \fso(W^\perp)$,
corresponding to a decomposition $V = W \oplus W^\perp$ where $W$ is
lorentzian and two-dimensional, corresponding to the string
worldsheet.  The odd subspace $\fk_1$ is isomorphic to the subspace
$\Delta_{\F1} \subset \Delta_+ \oplus \Delta_+$ given by
\begin{equation}
  \Delta_{\F1}   = \left\{ \begin{pmatrix} \varepsilon_1 \\
      \varepsilon_2 \end{pmatrix} \in \Delta_+ \oplus \Delta_+ \middle
    | \bnu_W  \begin{pmatrix} \varepsilon_1 \\
      \varepsilon_2 \end{pmatrix}  = \begin{pmatrix} - \varepsilon_1 \\
      \varepsilon_2 \end{pmatrix} \right\}~,
\end{equation}
where the Clifford endomorphism $\bnu_W$ corresponding to the volume
form of the string worldsheet is skewsymmetric relative to the spinor
inner product and obeys $\bnu_W^2 = +\1$.

Let $\be_\mu$ and $\be_a$ span $W$ and $W^\perp$, respectively, and
let $P_\mu$, $L_{\mu\nu}=\epsilon_{\mu\nu}L$ and $L_{ab}$ be the
generators of $\fk_0$.  Let $\beps_\alpha$ and $\bar\beps_\alphabar$
be basis elements for the subspaces of $\Delta_+$ satisfying $\bnu_W
\beps_\alpha = -\beps_\alpha$ and $\bnu_W \bar\beps_\alphabar =
\bar\beps_\alphabar$, respectively.\footnote{Here and in the sequel we
  use the bars on the spinors to distinguish them from the unbarred
  spinors and \emph{not} to denote the Dirac conjugate.}  Then
$\begin{pmatrix}\beps_\alpha\\0\end{pmatrix}$ and $\begin{pmatrix}0\\
  \bar\beps_\alphabar\end{pmatrix}$ span $\Delta_{\F1}$.  We let
$Q_\alpha$ and $\bar Q_\alphabar$ denote the corresponding basis for
$\fk_1$.  The nonzero Lie brackets in this basis are given, in
addition to those of $\fso(W^\perp)$, by
\begin{equation}
  \label{eq:IIBF1KSA}
  \begin{aligned}[m]
    [L,Q_\alpha] &= \half Q_\alpha\\
    [L,\bar Q_\alphabar] &=  - \half \bar Q_\alphabar\\
    [L,P_\mu] &=  \epsilon_\mu{}^\nu P_\nu\\
    [Q_\alpha, Q_\beta] &= \Gamma^\mu_{\alpha\beta} P_\mu\\
    [\bar Q_\alphabar, \bar Q_\betabar] &= \Gamma^\mu_{\alphabar\betabar} P_\mu~,
  \end{aligned}
\end{equation}
where
\begin{equation}
  \Gamma^\mu_{\alpha\beta} := \left<\beps_\alpha, \Gamma^\mu 
    \beps_\beta\right>
  \qquad\text{and}\qquad
  \Gamma^\mu_{\alphabar\betabar} := \left<\bar\beps_\alphabar, \Gamma^\mu 
    \bar\beps_\betabar\right>~.
\end{equation}

The ideal $I < \fk$ contains the supertranslation ideal and the
generator $L$, but $H^2(\fk;\fk)$ can be computed from the complex
$C^\bullet:=C^\bullet(I;\fk)^{\fr}$ of cochains which are invariant
under the reductive subalgebra $\fr:= \fso(W) \oplus \fso(W^\perp)$.
Letting $L^*$, $P^\mu$, $Q^\alpha$ and $\bar Q^\alphabar$ denote the
canonical dual basis for $I^*$, the differential $d$ in $C^\bullet$ is
determined uniquely by the following relations
\begin{equation}
  \begin{aligned}[m]
    d P^\mu &= \half \Gamma^\mu_{\alpha\beta} Q^\alpha \wedge Q^\beta
    + \half \Gamma^\mu_{\alphabar\betabar} \bar Q^\alphabar \wedge
    \bar Q^\betabar + \epsilon^\mu{}_\nu L^* \wedge P^\nu \\
    d Q^\alpha &= - \half L^* \wedge Q^\alpha\\
    d \bar Q^\alphabar &= \half L^* \wedge \bar Q^\alphabar\\
    d L^* &= 0\\
    d P_\mu &= L^* \otimes \epsilon_\mu{}^\nu P_\nu\\
    d Q_\alpha &= \half L^* \otimes Q_\alpha -
    \Gamma^\mu_{\alpha\beta} Q^\beta \otimes  P_\mu\\
    d \bar Q_\alphabar &= - \half L^* \otimes \bar Q_\alphabar -
    \Gamma^\mu_{\alphabar\betabar} \bar Q^\betabar \otimes  P_\mu\\
    d L &= - P^\mu \otimes \epsilon_\mu{}^\nu P_\nu + \half Q^\alpha
    \otimes Q_\alpha - \half \bar Q^\alphabar \otimes \bar Q_\alphabar\\
    d L_{ab} &= \half Q^\alpha \otimes \Gamma_{ab} \cdot Q_\alpha +
    \half \bar Q^\alphabar \otimes \Gamma_{ab}  \cdot \bar Q_\alphabar~.
  \end{aligned}
\end{equation}

The space of $0$-cochains is $1$-dimensional and spanned by $L$, but
since $dL \neq 0$, $H^0(\fk;\fk) = 0$ and $\dim B^1 = 1$.  The space
$C^1$ is $5$-dimensional and is spanned by $P^\mu \otimes P_\mu$,
$P^\mu \otimes \epsilon_\mu{}^\nu P_\nu$, $L^*\otimes L$, $Q^\alpha
\otimes Q_\alpha$ and $\bar Q^\alphabar \otimes \bar Q_\alphabar$.  As
in the case of the D-string, this can be understood by the fact that
the ideal $I$ is graded by the action of $2L$ with $L$ having degree
$0$, $Q_\alpha$ and $\bar Q_\alphabar$ having degrees $\pm 1$,
respectively, and $P_\mu$ having pieces of degrees $\pm 2$,
corresponding to a Witt basis for $W$.  The $5$-dimensional space of
cochains can be thought of as spanned by the cochains corresponding to
the identity maps of each of the five graded subspaces.  The
space $C^2$ of $2$-cochains is $7$-dimensional, spanned by
$P^\mu\wedge P^\nu\otimes \epsilon_{\mu\nu} L$, $L^* \wedge P^\mu
\otimes P_\mu$, $L^* \wedge P^\mu \otimes \epsilon_\mu{}^\nu P_\nu$,
$L^* \wedge Q^\alpha \otimes Q_\alpha$, $L^* \wedge \bar Q^\alphabar
\otimes \bar Q_\alphabar$, $Q^\alpha \wedge Q^\beta \otimes
\Gamma^\mu_{\alpha\beta} P_\mu$ and $\bar Q^\alphabar \wedge \bar Q^\betabar \otimes
\Gamma^\mu_{\alphabar\betabar} P_\mu$.

Computing the differential $d: C^1 \to C^2$, we find that
$H^1(\fk;\fk) \cong \RR$, with representative cocycle $\varphi:= 2
P^\mu \otimes P_\mu - Q^\alpha \otimes Q_\alpha - \bar Q^\alphabar
\otimes \bar Q_\alphabar$.  Similarly, computing $d:C^2 \to C^3$ we
find that $H^2(\fk;\fk)\cong \RR$, with representative cocycle $L^*
\wedge \varphi$.  This infinitesimal deformation integrates to a
one-parameter family of Lie superalgebras with brackets
\begin{equation}
  \label{eq:IIBF1deformed}
  \begin{aligned}[m]
    [L,Q_\alpha] &= (t + \half) Q_\alpha \\
    [L,\bar Q_\alphabar] &= (t - \half) \bar Q_\alphabar \\
    [L,P_\mu] &=  2 t P_\mu + \epsilon_\mu{}^\nu P_\nu\\
    [Q_\alpha, Q_\beta] &= \Gamma^\mu_{\alpha\beta} P_\mu\\
    [\bar Q_\alphabar, \bar Q_\betabar] &= \Gamma^\mu_{\alphabar\betabar} P_\mu
  \end{aligned}
\end{equation}
in addition to those involving $\fso(W^\perp)$ which remain
undeformed.  This deformation consists of changing the $L$-weight of
the generators in the Lie superalgebra in such a way that the $QQ$ and
$\bar Q\bar Q$ brackets remains invariant.  As in the case of the
D-string, this deformation is reminiscent of the construction of
two-dimensional topological conformal field theories via twisting.

\subsection{Rigidity of the D3-brane superalgebra}
\label{sec:D3}

The Killing superalgebra of the D3-brane background is the subalgebra
$\fk$ of the type IIB Poincaré superalgebra with $\fk_0 = \fso(W)
\oplus W \oplus \fso(W^\perp)$ where $V = W \oplus W^\perp$, with $W$
a four-dimensional lorentzian subspace.  The odd subspace $\fk_1$ is
isomorphic to the subspace $\Delta_{\D3} \subset \Delta_+ \oplus
\Delta_+$ defined by the graph of the endomorphism $\bnu_W : \Delta_+
\to \Delta_+$ corresponding to the volume form of $W$.  This
endomorphism obeys $\bnu_W^2 = - \1$ and, when extended to the
irreducible Clifford module, is symmetric with respect to the spinor
inner product.

Let $\be_\mu$ and $\be_a$ span $W$ and $W^\perp$, respectively and let
$\beps_\alpha$ span $\Delta_+$.  Let $\bpsi_\alpha =
\tfrac1{\sqrt{2}} \begin{pmatrix}\beps_\alpha\\\bnu_W 
  \beps_\alpha\end{pmatrix}$ be a basis for $\Delta_{\D3}$.  The
corresponding basis of $\fk$ is given by $P_\mu$, $L_{\mu\nu}$,
$L_{ab}$ and $Q_\alpha$.  The Lie brackets are inherited from those in
equation \eqref{eq:IIBPoincare} and are given explicitly, in addition
to those involving $L_{\mu\nu}$ and $L_{ab}$, by
\begin{equation}
  \label{eq:IIBD3KSA}
    [Q_\alpha, Q_\beta] = \Gamma^\mu_{\alpha\beta} P_\mu~,
\end{equation}
where
\begin{equation}
  \Gamma^\mu_{\alpha\beta} := \langle\bpsi_\alpha, \Gamma^\mu 
    \bpsi_\beta\rangle =  \left<\beps_\alpha, \Gamma^\mu 
    \beps_\beta\right>~.
\end{equation}

The ideal $I<\fk$ is spanned by $P_\mu$ and $Q_\alpha$ and the
semisimple factor $\fs$ by $L_{\mu\nu}$ and $L_{ab}$.  Letting $P^\mu$
and $Q^\alpha$ denote the canonical dual basis for $I^*$, the
differential $d$ on the complex $C^\bullet = C^\bullet(I;\fk)^{\fs}$
is determined by the following relations
\begin{equation}
  \begin{aligned}[m]
    d P^\mu &= \half \Gamma^\mu_{\alpha\beta} Q^\alpha \wedge Q^\beta\\
    d Q^\alpha &= 0\\
    d P_\mu &= 0\\
    d Q_\alpha &= - \Gamma^\mu_{\alpha\beta} Q^\beta \otimes  P_\mu\\
    d L_{\mu\nu} &= \eta_{\mu\rho} P^\rho \otimes P_\nu - \eta_{\nu\rho} P^\rho \otimes P_\mu + \half
    Q^\alpha \otimes \Gamma_{\mu\nu} \cdot Q_\alpha\\
    d L_{ab} &= \half Q^\alpha \otimes \Gamma_{ab} \cdot Q_\alpha~.
  \end{aligned}
\end{equation}

There are no nonzero $0$-cochains, since $C^0 = \fk^{\fs}$ and there
are no scalars in the superalgebra.  There is a $3$-dimensional space
of $1$-cochains, spanned by $P^\mu \otimes P_\mu$, $Q^\alpha \otimes
Q_\alpha$ and $Q^\alpha\otimes \bnu_W \cdot Q_\alpha$.  The first
cochain is the identity map $W \to W$, whereas the other two are
linear combinations involving the identity maps of the two irreducible
complex representations of $\fs$ into which the complexification of
$\Delta_{\D3}$ decomposes.  In terms of real maps, we have the identity
and the complex structure $\bnu_W$.  The space of $2$-cochains is also
$3$-dimensional, spanned by the cochains corresponding to the natural
isomorphism $\Lambda^2W \to \fso(W)$ and to its precomposition
with the Hodge star $\star: \Lambda^2W \to \Lambda^2W$,  as well as
to the projection $S^2 \Delta_{\D3} \to W$.  The corresponding cochains
are $P^\mu \wedge P^\nu \otimes L_{\mu\nu}$, $P^\mu \wedge P^\nu
\otimes \epsilon_{\mu\nu}{}^{\rho\sigma} L_{\rho\sigma}$ and $Q^\alpha
\wedge Q^\beta \otimes \Gamma^\mu_{\alpha\beta} P_\mu$.

Computing the differentials $d:C^1 \to C^2$ and $d:C^2 \to C^3$, we
find that $H^1(\fk;\fk) \cong \RR^2$ with representative cocycles $2
P^\mu \otimes P_\mu - Q^\alpha \otimes Q_\alpha$ and $Q^\alpha \otimes
\bnu_W \cdot Q_\alpha$.  The fact that this latter cochain is a
cocycle rests on the skewsymmetry of $(\Gamma^\mu
\bnu_W)_{\alpha\beta}$ in $\alpha\leftrightarrow\beta$.  Similarly, we
find that $H^2(\fk;\fk) = 0$, whence the D3-brane superalgebra is
rigid.

This rigidity might seem a little unexpected due to the fact that the
four-dimensional Poincaré superalgebra admits a deformation
\cite{ZuminoAdS,TripathyPatra}.  The calculation that the
anti-de~Sitter superalgebra is the unique deformation of the
four-dimensional Poincaré superalgebra had been announced in
\cite{Binegar}, but the expression of the deformed algebra in that
paper is incorrect.  The calculation in \cite{TripathyPatra} is
correct, but we find that the way of writing the algebra is perhaps
not the most transparent.  For this reason we present this calculation
in the following section.

\subsubsection{A deformation of the four-dimensional Poincaré superalgebra}
\label{sec:4d}

We choose to do the calculation using two-component spinor language,
which simplifies many of the calculations.  Our conventions are taken
from \cite[Appendix~B]{JMFBUSSTEPP}.  The generators of the Poincaré
superalgebra $\fp$ are $Q_\alpha$, $\bar Q_\alphadot$,
$P_{\alpha\alphadot}$, $L_{\alpha\beta}$ and $\bar
L_{\alphadot\betadot}$, with Lie brackets
\begin{equation}
  \label{eq:4dPoincare}
  \begin{aligned}[m]
    [L_{\alpha\beta},Q_\gamma] &= \epsilon_{\beta\gamma} Q_\alpha + \epsilon_{\alpha\gamma} Q_\beta \\
    [L_{\alpha\beta},P_{\gamma\alphadot}] &= \epsilon_{\beta\gamma}
    P_{\alpha\alphadot} + \epsilon_{\alpha\gamma} P_{\beta\alphadot}\\  
    [Q_\alpha, \bar Q_\betadot] &= P_{\alpha\alphadot}~,
  \end{aligned}
\end{equation}
together with the conjugate versions of the first two brackets,
obtained from those by the replacements $\epsilon_{\alpha\beta}
\mapsto \bar\epsilon_{\alphadot\betadot}$, $L_{\alpha\beta} \mapsto
\bar L_{\alphadot\betadot}$, $P_{\alpha\betadot} \mapsto
P_{\beta\alphadot}$ and $Q_\alpha \mapsto \bar Q_\alphadot$.

Let $I$ be the ideal spanned by $P_{\alpha\alphadot}$, $Q_\alpha$ and
$\bar Q_\alphadot$ and $\fs$ the semisimple factor spanned by
$L_{\alpha\beta}$ and $\bar L_{\alphadot\betadot}$.  We denote by
$P^{\alpha\alphadot}$, $Q^\alpha$ and $\bar Q^{\alphadot}$ the
canonical dual basis for $I^*$.  The differential in the complex
$C^\bullet := C^\bullet(I;\fp)^{\fs}$ is determined by the following
relations
\begin{equation}
  \begin{aligned}[m]
    d P^{\alpha\alphadot} &= Q^\alpha \wedge \bar Q^\alphadot\\
    d Q^\alpha &= 0\\
    d P_{\alpha\alphadot} &= 0\\
    d Q_\alpha &= - \bar Q^\alphadot \otimes  P_{\alpha\alphadot}\\
    d L_{\alpha\beta} &= - \epsilon_{\alpha\gamma} P^{\gamma\gammadot}
    \otimes P_{\beta\gammadot} - \epsilon_{\beta\gamma}
    P^{\gamma\gammadot} \otimes P_{\alpha\gammadot} +
    \epsilon_{\alpha\gamma} Q^\gamma\otimes Q_\beta +
    \epsilon_{\beta\gamma} Q^\gamma\otimes Q_\alpha~,
  \end{aligned}
\end{equation}
and their conjugates.

There are no $0$-cochains since there are no Lorentz scalars in the
algebra.  The space of $1$-cochains is three-dimensional, spanned by
$Q^\alpha \otimes Q_\alpha$, $\bar Q^\alphadot \otimes \bar
Q_\alphadot$ and $P^{\alpha\alphadot}\otimes P_{\alpha\alphadot}$.
The space of $2$-cochains is $7$-dimensional and spanned by
$P^{\alpha\alphadot} \wedge P^{\beta\betadot} \otimes
\epsilon_{\alpha\beta} \bar L_{\alphadot\betadot}$,
$P^{\alpha\alphadot} \wedge P^{\beta\betadot} \otimes
\bar \epsilon_{\alphadot\betadot} L_{\alpha\beta}$,
$P^{\alpha\alphadot} \wedge Q^\beta \otimes \epsilon_{\alpha\beta}
\bar Q_{\alphadot}$, 
$P^{\alpha\alphadot} \wedge \bar Q^\betadot \otimes \bar \epsilon_{\alphadot\betadot}
Q_{\alpha}$, $Q^\alpha \wedge \bar Q^{\alphadot} \otimes
P_{\alpha\alphadot}$, $Q^\alpha \wedge Q^\beta \otimes
L_{\alpha\beta}$ and $\bar Q^\alphadot \wedge \bar Q^\betadot \otimes
\bar L_{\alphadot\betadot}$.

Computing the differential $d:C^1 \to C^2$ we find a two-dimensional
space of cocycles, spanned by $2P^{\alpha\alphadot}\otimes
P_{\alpha\alphadot} - Q^\alpha \otimes Q_\alpha - \bar Q^\alphadot \otimes \bar
Q_\alphadot$ and $Q^\alpha \otimes Q_\alpha - \bar Q^\alphadot \otimes \bar
Q_\alphadot$.  Strictly speaking the latter cocycle is not real, so we
would have to multiply by $i$ in order to make it real.  It
corresponds to the map on spinors induced by multiplication with the
volume form; that is, $\gamma_5$ in old money.  Computing the
differential $d:C^2 \to C^3$ we also find a two-dimensional space of
cocycles, spanned by $2 P^{\alpha\alphadot} \wedge Q^\beta \otimes
\epsilon_{\alpha\beta} \bar Q_{\alphadot} + Q^\alpha \wedge Q^\beta \otimes
L_{\alpha\beta}$ and its conjugate.  Since we are interested in
deformations of the real form of the Lie superalgebra, we choose the
real part of the cocycle; that is,
\begin{multline}
  2 P^{\alpha\alphadot} \wedge Q^\beta \otimes
\epsilon_{\alpha\beta} \bar Q_{\alphadot} +
  2 P^{\alpha\alphadot} \wedge \bar Q^\betadot \otimes
\bar\epsilon_{\alphadot\betadot} Q_{\alpha} + Q^\alpha \wedge Q^\beta \otimes
L_{\alpha\beta} + \bar Q^\alphadot \wedge \bar Q^\betadot \otimes \bar
L_{\alphadot\betadot}~,
\end{multline}
corresponding to the following Lie brackets to first order in the
deformation parameter $t$:
\begin{equation}
  \begin{aligned}[m]
    [P_{\alpha\alphadot},Q_\beta] &= - t \epsilon_{\alpha\beta} \bar
    Q_{\alphadot}\\
    [Q_\alpha, Q_\beta] &= -t L_{\alpha\beta}\\
    [Q_\alpha, \bar Q_\betadot] &= P_{\alpha\betadot}~,
  \end{aligned}
\end{equation}
and their conjugates.  There is an obstruction to integrating this
deformation at the next order, which requires introducing the bracket
\begin{equation}
  [P_{\alpha\alphadot}, P_{\beta\betadot}] = t^2 
  \epsilon_{\alpha\beta} \bar L_{\alphadot\betadot} +  t^2 \bar
  \epsilon_{\alphadot\betadot} L_{\alpha\beta}~.
\end{equation}
The above brackets, together with the ones involving the Lorentz
generators, define a one-parameter family of deformations, first
written down in \cite{ZuminoAdS}, corresponding to the $AdS_4$
superalgebra.  The algebraic reason why this deformation of the
$4$-dimensional Poincaré superalgebra does not lift to a deformation
of the D3-brane superalgebra is that the ten-dimensional chirality of
the IIB spinors forbids the necessary extra terms in the $QQ$ bracket.

\subsection{Rigidity of the D5-brane superalgebra}
\label{sec:IIBD5}

The Killing superalgebra of the type IIB D5-brane is the
subsuperalgebra $\fk = \fk_0 \oplus \fk_1$ of the type IIB Poincaré
superalgebra with $\fk_0 = \fso(W) \oplus W \oplus \fso(W^\perp)$,
where $V = W \oplus W^\perp$ and $W$ a six-dimensional lorentzian
subspace, and $\fk_1$ isomorphic to the subspace $\Delta_{\D5} \subset
\Delta_+ \oplus \Delta_+$ defined as the graph of the volume form
$\bnu_W : \Delta_+ \oplus \Delta_+$, which obeys $\bnu_W^2 = +\1$ and
is skewsymmetric relative to the spinor inner product when extended to
a Clifford endomorphism.

Let $\be_\mu$ and $\be_a$ span $W$ and $W^\perp$, respectively and let
$\beps_\alpha$ span $\Delta_+$.  Let $\bpsi_\alpha =
\tfrac1{\sqrt{2}} \begin{pmatrix}\beps_\alpha\\\bnu_W 
  \beps_\alpha\end{pmatrix}$ be a basis for $\Delta_{\D5}$.  The 
corresponding basis of $\fk$ is given by $P_\mu$, $L_{\mu\nu}$,
$L_{ab}$ and $Q_\alpha$.  The Lie brackets are inherited from those in
equation \eqref{eq:IIBPoincare} and are given explicitly, in addition
to those involving $L_{\mu\nu}$ and $L_{ab}$, by
\begin{equation}
  \label{eq:IIBD5KSA}
    [Q_\alpha, Q_\beta] = \Gamma^\mu_{\alpha\beta} P_\mu~,
\end{equation}
where
\begin{equation}
  \Gamma^\mu_{\alpha\beta} := \langle\bpsi_\alpha, \Gamma^\mu 
    \bpsi_\beta\rangle =  \left<\beps_\alpha, \Gamma^\mu 
    \beps_\beta\right>~.
\end{equation}

The ideal $I<\fk$ is spanned by $P_\mu$ and $Q_\alpha$ and the
semisimple factor $\fs$ by $L_{\mu\nu}$ and $L_{ab}$.  Letting $P^\mu$
and $Q^\alpha$ denote the canonical dual basis for $I^*$, the
differential $d$ on the complex $C^\bullet = C^\bullet(I;\fk)^{\fs}$
is determined by the following relations
\begin{equation}
  \begin{aligned}[m]
    d P^\mu &= \half \Gamma^\mu_{\alpha\beta} Q^\alpha \wedge Q^\beta\\
    d Q^\alpha &= 0\\
    d P_\mu &= 0\\
    d Q_\alpha &= - \Gamma^\mu_{\alpha\beta} Q^\beta \otimes  P_\mu\\
    d L_{\mu\nu} &= \eta_{\mu\rho} P^\rho \otimes P_\nu -
    \eta_{\nu\rho} P^\rho \otimes P_\mu + \half Q^\alpha \otimes
    \Gamma_{\mu\nu} \cdot Q_\alpha\\
    d L_{ab} &= \half Q^\alpha \otimes \Gamma_{ab} \cdot Q_\alpha~.
  \end{aligned}
\end{equation}

There are no nonzero $0$-cochains, since $C^0 = \fk^{\fs}$ and there
are no scalars in the superalgebra.  There is a $3$-dimensional space
of $1$-cochains, spanned by $P^\mu \otimes P_\mu$, $Q^\alpha \otimes
Q_\alpha$ and $Q^\alpha\otimes \bnu_W \cdot Q_\alpha$.  The space of
$2$-cochains is also $3$-dimensional, spanned by the following
cochains: $P^\mu \wedge P^\nu \otimes L_{\mu\nu}$, $Q^\alpha \wedge
Q^\beta \otimes \Gamma^\mu_{\alpha\beta} P_\mu$ and $Q^\alpha \wedge
Q^\beta \otimes \left(\Gamma^\mu\bnu_W\right)_{\alpha\beta} P_\mu$.

Computing the differentials $d:C^1 \to C^2$ and $d:C^2 \to C^3$, we
find that $H^1(\fk;\fk) \cong \RR$ with representative cocycle $2
P^\mu \otimes P_\mu - Q^\alpha \otimes Q_\alpha$, whence $
\dim B^2 = 2$.  Since $d:C^2 \to C^3$ is not identically zero, we
conclude that $H^2(\fk;\fk) = 0$, whence the D5-brane superalgebra is
rigid.

\subsection{Rigidity of the NS 5-brane superalgebra}
\label{sec:IIBNS5}

The Killing superalgebra of the type IIB NS5-brane is the
subsuperalgebra $\fk = \fk_0 \oplus \fk_1$ of the IIB Poincaré 
superalgebra where $\fk_0 = \fso(W) \oplus W \oplus \fso(W^\perp)$,
corresponding to a decomposition $V = W \oplus W^\perp$ where $W$ is
lorentzian and six-dimensional, corresponding to the brane
worldvolume.  The odd subspace $\fk_1$ is isomorphic to the subspace
$\Delta_{\NS5} \subset \Delta_+ \oplus \Delta_+$ given by
\begin{equation}
  \Delta_{\NS5}   = \left\{ \begin{pmatrix} \varepsilon_1 \\
      \varepsilon_2 \end{pmatrix} \in \Delta_+ \oplus \Delta_+ \middle
    | \bnu_W  \begin{pmatrix} \varepsilon_1 \\
      \varepsilon_2 \end{pmatrix}  = \begin{pmatrix} - \varepsilon_1 \\
      \varepsilon_2 \end{pmatrix} \right\}~,
\end{equation}
where the Clifford endomorphism $\bnu_W$ corresponding to the volume
form of the brane worldvolume is skewsymmetric relative to the spinor
inner product and obeys $\bnu_W^2 = +\1$.

Let $\be_\mu$ and $\be_a$ span $W$ and $W^\perp$, respectively, and
let $P_\mu$, $L_{\mu\nu}$ and $L_{ab}$ be the generators of $\fk_0$.
Let $\beps_\alpha$ and $\bar\beps_\alphabar$ be basis elements for the
subspaces of $\Delta_+$ satisfying $\bnu_W  \beps_\alpha =
-\beps_\alpha$ and $\bnu_W  \bar\beps_\alphabar =
\bar\beps_\alphabar$, respectively, so that
$\begin{pmatrix}\beps_\alpha\\0\end{pmatrix}$ and $\begin{pmatrix}0\\
  \bar\beps_\alphabar\end{pmatrix}$ span $\Delta_{\F1}$.  We let $Q_\alpha$
and $\bar Q_\alphabar$ denote the corresponding basis for $\fk_1$.
The nonzero Lie brackets in this basis are given, in addition to those
of $\fk_0$, by
\begin{equation}
  \label{eq:IIBNS5KSA}
  \begin{aligned}[m]
    [L_{\mu\nu},Q_\alpha] &= \half \Gamma_{\mu\nu} \cdot Q_\alpha\\
    [L_{\mu\nu},\bar Q_\alphabar] &= \half \Gamma_{\mu\nu} \cdot \bar Q_\alphabar\\
    [L_{ab},Q_\alpha] &= \half \Gamma_{ab} \cdot Q_\alpha\\
    [L_{ab},\bar Q_\alphabar] &= \half \Gamma_{ab} \cdot  \bar Q_\alphabar\\
    [Q_\alpha, Q_\beta] &= \Gamma^\mu_{\alpha\beta} P_\mu\\
    [\bar Q_\alphabar, \bar Q_\betabar] &= \Gamma^\mu_{\alphabar\betabar} P_\mu~,
  \end{aligned}
\end{equation}
where, as before,
\begin{equation}
  \Gamma^\mu_{\alpha\beta} := \left<\beps_\alpha, \Gamma^\mu 
    \beps_\beta\right>
  \qquad\text{and}\qquad
  \Gamma^\mu_{\alphabar\betabar} := \left<\bar\beps_\alphabar, \Gamma^\mu 
    \bar\beps_\betabar\right>~.
\end{equation}

The ideal $I < \fk$ is spanned by $P_\mu$, $Q_\alpha$ and $\bar
Q_\alphabar$ and the differential in the complex
$C^\bullet:=C^\bullet(I;\fk)^{\fs}$ of cochains which are invariant
under the semisimple subalgebra $\fs:= \fso(W) \oplus \fso(W^\perp)$
is determined uniquely by the following relations
\begin{equation}
  \begin{aligned}[m]
    d P^\mu &= \half \Gamma^\mu_{\alpha\beta} Q^\alpha \wedge Q^\beta
    + \half \Gamma^\mu_{\alphabar\betabar} \bar Q^\alphabar \wedge
    \bar Q^\betabar\\
    d Q^\alpha &= 0\\
    d \bar Q^\alphabar &= 0\\
    d P_\mu &= 0\\
    d Q_\alpha &= - \Gamma^\mu_{\alpha\beta} Q^\beta \otimes  P_\mu\\
    d \bar Q_\alphabar &= - \Gamma^\mu_{\alphabar\betabar} \bar
    Q^\betabar \otimes  P_\mu\\
    d L_{\mu\nu} &=  \eta_{\mu\rho} P^\rho \otimes P_\nu -
    \eta_{\nu\rho} P^\rho \otimes P_\mu + \half Q^\alpha \otimes
    \Gamma_{\mu\nu} \cdot Q_\alpha + \half \bar Q^\alphabar \otimes
    \Gamma_{\mu\nu} \cdot \bar Q_\alphabar\\
    d L_{ab} &= \half Q^\alpha \otimes \Gamma_{ab} \cdot Q_\alpha +
    \half \bar Q^\alphabar \otimes \Gamma_{ab}  \cdot \bar Q_\alphabar~.
  \end{aligned}
\end{equation}

There are no $0$-cochains.  The space $C^1$ is $3$-dimensional and is
spanned by the cochains corresponding to the identity maps $P^\mu
\otimes P_\mu$, $Q^\alpha \otimes Q_\alpha$ and $\bar Q^\alphabar
\otimes \bar Q_\alphabar$.  The space $C^2$ of $2$-cochains is
$3$-dimensional, spanned by $P^\mu\wedge P^\nu\otimes L_{\mu\nu}$,
$Q^\alpha \wedge Q^\beta \otimes \Gamma^\mu_{\alpha\beta} P_\mu$ and
$\bar Q^\alphabar \wedge \bar Q^\betabar \otimes
\Gamma^\mu_{\alphabar\betabar} P_\mu$.

Computing the differential $d: C^1 \to C^2$, we find that
$H^1(\fk;\fk) \cong \RR$, with representative cocycle $\varphi:= 2
P^\mu \otimes P_\mu - Q^\alpha \otimes Q_\alpha - \bar Q^\alphabar
\otimes \bar Q_\alphabar$.  Since $d:C^2 \to C^3$ is not the zero map,
we conclude that $H^2(\fk;\fk) = 0$, proving the rigidity of the IIB
NS5 superalgebra.

\subsection{A deformation of the D7-brane superalgebra}
\label{sec:D7}

The Killing superalgebra of the D7-brane is the
subsuperalgebra $\fk$ of the IIB Poincaré superalgebra with $\fk_0 =
\fso(W) \oplus W \oplus \fso(W^\perp)$, where $V = W \oplus W^\perp$
is the decomposition of the $10$-dimensional lorentzian vector space
into an $8$-dimensional lorentzian subspace $W$, corresponding to the
brane worldvolume and its $2$-dimensional perpendicular complement.
The odd subspace $\fk_1$ is isomorphic to the graph $\Delta_{\D7}
\subset \Delta_+ \oplus \Delta_+$ of the endomorphism $\bnu_W:
\Delta_+ \to \Delta_+$ corresponding to the volume form of $W$.  As in
the D3-brane, the extension of $\bnu_W$ to the Clifford module
is symmetric relative to the spinor inner product and obeys
$\bnu_W^2 = - \1$.

Let $\be_\mu$ and $\be_a$ span $W$ and $W^\perp$, respectively and let
$\beps_\alpha$ span $\Delta_+$.  Let $\bpsi_\alpha =
\tfrac1{\sqrt{2}} \begin{pmatrix}\beps_\alpha\\\bnu_W 
  \beps_\alpha\end{pmatrix}$ be a basis for $\Delta_{\D7}$.  The
corresponding basis of $\fk$ is given by $P_\mu$, $L_{\mu\nu}$,
$L_{ab} = \epsilon_{ab}L$ and $Q_\alpha$.  The Lie brackets are
inherited from those in equation \eqref{eq:IIBPoincare} and are given
explicitly, in addition to those involving the Lorentz subalgebra of
the brane worldvolume, by
\begin{equation}
  \label{eq:IIBD7KSA}
  \begin{aligned}[m]
    [L,Q_\alpha] &= - \half \bnu_W \cdot Q_\alpha\\
    [Q_\alpha, Q_\beta] &= \Gamma^\mu_{\alpha\beta} P_\mu~,
  \end{aligned}
\end{equation}
where
\begin{equation}
  \Gamma^\mu_{\alpha\beta} := \langle\bpsi_\alpha, \Gamma^\mu 
    \bpsi_\beta\rangle =  \left<\beps_\alpha, \Gamma^\mu 
    \beps_\beta\right>~.
\end{equation}

The ideal $I < \fk$ includes the generator $L$, but we may work with
cochains which are invariant under the reductive subalgebra $\fr :=
\fso(W) \oplus \fso(W^\perp)$.  Letting $L^*$, $P^\mu$ and $Q^\alpha$
be a basis for $I^*$, the differential in the complex $C^\bullet =
C^\bullet(I,\fk)^{\fr}$ is determined by the following relations
\begin{equation}
  \begin{aligned}[m]
    d P^\mu &= \half \Gamma^\mu_{\alpha\beta} Q^\alpha \wedge Q^\beta\\
    d Q^\alpha &= \half L^* \wedge \left(\bnu_W\right)^\alpha{}_\beta Q^\beta\\
    d L^* &= 0\\
    d P_\mu &= 0\\
    d Q_\alpha &= -\half L^* \otimes \bnu_W \cdot Q_\alpha -
    \Gamma^\mu_{\alpha\beta} Q^\beta \otimes  P_\mu\\
    d L &= - \half Q^\alpha \otimes \bnu_W \cdot Q_\alpha\\
    d L_{\mu\nu} &= \eta_{\mu\rho} P^\rho \otimes P_\nu -
    \eta_{\nu\rho} P^\rho \otimes P_\mu + \half Q^\alpha \otimes
    \Gamma_{\mu\nu} \cdot Q_\alpha~.
  \end{aligned}
\end{equation}

The $0$-cochains $C^0 = \fk^{\fr}$ are spanned by $L$, but $dL \neq
0$, hence $H^0(\fk;\fk) = 0$ and $\dim B^1 = 1$.  The space of
$1$-cochains is $4$-dimensional, spanned by $P^\mu \otimes P_\mu$,
$L^*\otimes L$, $Q^\alpha \otimes Q_\alpha$ and $Q^\alpha \otimes
\bnu_W \cdot Q_\alpha$.  The space of $2$-cochains is $5$-dimensional,
spanned by $P^\mu\wedge P^\nu\otimes L_{\mu\nu}$, $L^* \wedge
P^\mu \otimes P_\mu$, $L^* \wedge Q^\alpha \otimes Q_\alpha$, $L^* \wedge Q^\alpha
\otimes \bnu_W \cdot Q_\alpha$ and $Q^\alpha \wedge Q^\beta \otimes
\Gamma^\mu_{\alpha\beta} P_\mu$.

Computing the differential $d: C^1 \to C^2$, we find that
$H^1(\fk;\fk) \cong \RR$, with representative cocycle $\varphi:= 2
P^\mu \otimes P_\mu - Q^\alpha \otimes Q_\alpha$.  Similarly,
computing $d:C^2 \to C^3$ we find that $H^2(\fk;\fk)\cong \RR$, with
representative cocycle $L^* \wedge \varphi$.  This infinitesimal
deformation integrates to a one-parameter family of Lie superalgebras
with brackets
\begin{equation}
  \label{eq:IIBD7deformed}
  \begin{aligned}[m]
    [L,Q_\alpha] &= t Q_\alpha - \half \bnu_W \cdot Q_\alpha\\
    [L,P_\mu] &=  2 t P_\mu\\
    [Q_\alpha, Q_\beta] &= \Gamma^\mu_{\alpha\beta} P_\mu
  \end{aligned}
\end{equation}
in addition to those involving $\fso(W)$, which remain undeformed.

\section{Type IIA backgrounds}
\label{sec:TypeIIA}

In this section we explore the Lie superalgebra deformations of the
Killing superalgebras of certain type IIA backgrounds.  We start with
the Minkowski vacuum and then go on to the elementary brane
backgrounds.

\subsection{Rigidity of the Poincaré superalgebra}
\label{sec:PoincareIIA}

The Killing superalgebra $\fk$ of the unique maximally supersymmetric
solution of type IIA supergravity is the IIA Poincaré superalgebra,
which extends the ten-dimensional Poincaré algebra $\fso(V) \oplus V$
by supercharges transforming in the spinorial representation $\Delta_+
\oplus \Delta_-$ of $\fso(V)$.  We will let $\be_\mu$ denote an
orthonormal basis for $V$ and $P_\mu$ and $L_{\mu\nu}$ the
corresponding basis for the Poincaré algebra.  We will let
$\beps_\alpha$ and $\bar \beps_\alphabar$ be a basis for $\Delta_\pm$,
respectively, and $Q_\alpha$ and $\bar Q_\alphabar$ the corresponding
basis for the odd subspace of the Poincaré superalgebra.  In this
basis, the nonzero Lie brackets are, in addition to those of the
Poincaré algebra, the following:
\begin{equation}
  \label{eq:IIAPoincare}
  \begin{aligned}[m]
    [L_{\mu\nu},Q_\alpha] &= \half \Gamma_{\mu\nu} \cdot Q_\alpha\\
    [L_{\mu\nu},\bar Q_\alphabar] &= \half \Gamma_{\mu\nu} \cdot \bar Q_\alphabar\\
    [Q_\alpha, Q_\beta] &= \Gamma^\mu_{\alpha\beta} P_\mu\\
    [\bar Q_\alphabar, \bar Q_\betabar] &= \Gamma^\mu_{\alphabar\betabar} P_\mu~,
  \end{aligned}
\end{equation}
where, as usual,
\begin{equation}
  \Gamma^\mu_{\alpha\beta} := \left<\beps_\alpha, \Gamma^\mu 
    \beps_\beta\right>
  \qquad\text{and}\qquad
  \Gamma^\mu_{\alphabar\betabar} := \left<\bar\beps_\alphabar, \Gamma^\mu 
    \bar\beps_\betabar\right>~.
\end{equation}

The supertranslation ideal is spanned by $P_\mu$, $Q_\alpha$ and $\bar
Q_\alphabar$ and the semisimple algebra is the Lorentz subalgebra $\fs
= \fso(V)$.  The complex computing the deformations is $C^\bullet =
C^\bullet(I;\fk)^{\fs}$ and its differential is determined uniquely
by its action on the above basis for $\fk$ and the canonical dual basis
$P^\mu$, $Q^\alpha$ and $\bar Q^\alphabar$ for $I^*$:
\begin{equation}
  \begin{aligned}[m]
    d P^\mu &= \half \Gamma^\mu_{\alpha\beta} Q^\alpha \wedge Q^\beta
    + \half \Gamma^\mu_{\alphabar\betabar} \bar Q^\alphabar \wedge
    \bar Q^\betabar\\
    d Q^\alpha &= 0\\
    d \bar Q^\alphabar &= 0\\
    d P_\mu &= 0\\
    d Q_\alpha &= - \Gamma^\mu_{\alpha\beta} Q^\beta \otimes  P_\mu\\
    d \bar Q_\alphabar &= - \Gamma^\mu_{\alphabar\betabar} \bar
    Q^\betabar \otimes  P_\mu\\
    d L_{\mu\nu} &=  \eta_{\mu\rho} P^\rho \otimes P_\nu -
    \eta_{\nu\rho} P^\rho \otimes P_\mu + \half Q^\alpha \otimes
    \Gamma_{\mu\nu} \cdot Q_\alpha + \half \bar Q^\alphabar \otimes
    \Gamma_{\mu\nu} \cdot \bar Q_\alphabar~.
  \end{aligned}
\end{equation}

There are no $0$-cochains and the space of $1$-cochains is
$3$-dimensional, spanned by $P^\mu \otimes P_\mu$, $Q^\alpha \otimes
Q_\alpha$ and $\bar Q^\alphabar \otimes \bar Q_\alphabar$,
corresponding to the identity maps $V \to V$ and $\Delta_\pm \to
\Delta_\pm$.  The space $C^2$ of $2$-cochains is $6$-dimensional,
spanned by $P^\mu\wedge P^\nu\otimes L_{\mu\nu}$, $P^\mu \wedge
Q^\alpha \otimes \Gamma_\mu \cdot Q_\alpha$, $P^\mu \wedge \bar
Q^\alphabar \otimes \Gamma_\mu  \cdot \bar Q_\alphabar$, $Q^\alpha
\wedge Q^\beta \otimes \Gamma^\mu_{\alpha\beta} P_\mu$, $\bar
Q^\alphabar \wedge \bar Q^\betabar \otimes
\Gamma^\mu_{\alphabar\betabar} P_\mu$ and $Q^\alpha \wedge \bar
Q^\betabar \otimes \Gamma^{\mu\nu}_{\alpha\betabar} L_{\mu\nu}$.
These cochains correspond to the isomorphism $\Lambda^2 V \to
\fso(V)$, Clifford multiplication $V \otimes \Delta_\pm \to
\Delta_\mp$ and the projections $S^2 \Delta_\pm \to V$ and $\Delta_+
\otimes \Delta_- \to \Lambda^2 V$.

Computing the differential $d: C^1 \to C^2$, we find
\begin{equation}
  \begin{aligned}[m]
    d \left(P^\mu \otimes P_\mu\right) &= \half
    \Gamma^\mu_{\alpha\beta} Q^\alpha \wedge Q^\beta \otimes  P_\mu +
    \half \Gamma^\mu_{\alphabar\betabar} \bar Q^\alphabar \wedge \bar
    Q^\betabar \otimes  P_\mu\\
    d \left(Q^\alpha \otimes Q_\alpha\right) &=
    \Gamma^\mu_{\alpha\beta} Q^\alpha \wedge Q^\beta \otimes P_\mu\\
    d \left(\bar Q^\alphabar \otimes \bar Q_\alphabar \right) &=
    \Gamma^\mu_{\alphabar\betabar} \bar Q^\alphabar \wedge \bar Q^\betabar \otimes P_\mu~,
  \end{aligned}
\end{equation}
so that $H^1(\fk;\fk) \cong \RR$, with representative cocycle $2 P^\mu
\otimes P_\mu - Q^\alpha \otimes Q_\alpha - \bar Q^\alphabar \otimes
\bar Q_\alphabar$.  This implies that $\dim B^2 = 2$, spanned by $d
(Q^\alpha \otimes Q_\alpha)$ and $d(\bar Q^\alphabar\otimes \bar
Q_\alphabar)$, say.

Computing the differential $d: C^2 \to C^3$, we find, in addition to
the coboundaries,
\begin{equation}
  \begin{aligned}[m]
    d \left( P^\mu\wedge P^\nu \otimes L_{\mu\nu} \right) &=
    \Gamma^\mu_{\alpha\beta} P^\nu \wedge Q^\alpha \wedge Q^\beta
  \otimes L_{\mu\nu} + \Gamma^\mu_{\alphabar\betabar} P^\nu \wedge
  \bar Q^\alphabar \wedge \bar Q^\betabar \otimes L_{\mu\nu}\\
  & \quad {} + \half P^\mu\wedge P^\nu \wedge Q^\alpha \otimes
  \Gamma_{\mu\nu} \cdot Q_\alpha + \half P^\mu\wedge P^\nu \wedge \bar
  Q^\alphabar \otimes \Gamma_{\mu\nu} \cdot  \bar Q_\alphabar\\
  d \left( P^\mu \wedge Q^\alpha \otimes \Gamma_\mu \cdot Q_\alpha
  \right) &= \half \Gamma^\mu_{\beta\gamma} Q^\beta \wedge Q^\gamma
  \wedge Q^\alpha \otimes \Gamma_\mu \cdot Q_\alpha + \half
  \Gamma^\mu_{\betabar\gammabar} \bar Q^\betabar \wedge \bar
  Q^\gammabar \wedge Q^\alpha \otimes \Gamma_\mu \cdot Q_\alpha\\
  &\quad {} - \left(\Gamma_\mu\right)^\betabar{}_\alpha
  \Gamma^\nu_{\betabar\gammabar} P^\mu \wedge Q^\alpha \wedge \bar
  Q^\gammabar \otimes P_\nu\\
  d \left( P^\mu \wedge \bar Q^\alphabar \otimes \Gamma_\mu  \bar
    Q_\alphabar\right) &= \half \Gamma^\mu_{\beta\gamma} Q^\beta \wedge Q^\gamma
  \wedge \bar Q^\alphabar \otimes \Gamma_\mu  \cdot \bar Q_\alphabar + \half
  \Gamma^\mu_{\betabar\gammabar} \bar Q^\betabar \wedge \bar
  Q^\gammabar \wedge \bar Q^\alphabar \otimes \Gamma_\mu  \cdot \bar Q_\alphabar\\
  &\quad {} - \left(\Gamma_\mu\right)^\beta{}_\alphabar
  \Gamma^\nu_{\beta\gamma} P^\mu \wedge \bar Q^\alphabar \wedge
  Q^\gamma \otimes P_\nu \\
  d \left( Q^\alpha \wedge \bar Q^\betabar \otimes
    \Gamma^{\mu\nu}_{\alpha\betabar} L_{\mu\nu}\right) &= 
  2 \eta_{\mu\rho} \Gamma^{\mu\nu}_{\alpha\betabar} Q^\alpha \wedge
  \bar Q^\betabar \wedge P^\rho \otimes P_\nu + \half 
  \Gamma^{\mu\nu}_{\alpha\betabar} Q^\alpha \wedge
  \bar Q^\betabar \wedge Q^\gamma \otimes \Gamma_{\mu\nu}  \cdot Q_\gamma \\
  & \quad {} + \half \Gamma^{\mu\nu}_{\alpha\betabar} Q^\alpha \wedge
  \bar Q^\betabar \wedge \bar Q^\gammabar \otimes \Gamma_{\mu\nu} \cdot
  \bar Q_\gammabar~.
  \end{aligned}
\end{equation}
It is clear that any possible cocycle must be a linear combination of
the last three cochains, since the differential of the first cochain
is the only one having explicit dependence on $L_{\mu\nu}$.
Therefore let
\begin{equation}
  \Theta = a_1 P^\mu \wedge Q^\alpha \otimes \Gamma_\mu \cdot Q_\alpha
  + a_2 P^\mu \wedge \bar Q^\alphabar \otimes \Gamma_\mu  \cdot \bar
  Q_\alphabar + a_3 Q^\alpha \wedge \bar Q^\betabar \otimes
  \Gamma^{\mu\nu}_{\alpha\betabar} L_{\mu\nu}
\end{equation}
and consider the equation $d\Theta=0$.  The coefficient of the
monomial $P^\mu \wedge Q^\alpha \wedge \bar Q^\betabar \otimes P_\nu$
is given by
\begin{equation}
  \label{eq:dT1}
  -a_1 \left(\Gamma_\mu\right)^\gammabar{}_\alpha
  \Gamma^\nu_{\betabar\gammabar} 
  -a_2 \left(\Gamma_\mu\right)^\gamma{}_\betabar
  \Gamma^\nu_{\gamma\alpha} + 2 a_3
  \left(\Gamma_\mu{}^\nu\right)_{\alpha\betabar}~.
\end{equation}
Now,
\begin{equation}
  \begin{split}
    \left(\Gamma_\mu\right)^\gammabar{}_\alpha
    \Gamma^\nu_{\betabar\gammabar} &= \langle\bar\beps_\betabar,
      \Gamma^\nu  \Gamma_\mu  \beps_\alpha\rangle\\
    &= - \left<\beps_\alpha, \Gamma_\mu  \Gamma^\nu  \bar
      \beps_\gammabar \right>\\
    &= - \left(\Gamma_\mu{}^\nu\right)_{\alpha\betabar} -
    \delta_\mu^\nu \left<\beps_\alpha, \bar\beps_\gammabar\right>~,
  \end{split}
\end{equation}
and similarly
\begin{equation}
  \left(\Gamma_\mu\right)^\gamma{}_\betabar
  \Gamma^\nu_{\gamma\alpha}  = - \left(\Gamma_\mu{}^\nu\right)_{\alpha\betabar} +
    \delta_\mu^\nu \left<\beps_\alpha, \bar\beps_\gammabar\right>~,
\end{equation}
whence the expression in equation~\eqref{eq:dT1} becomes
\begin{equation}
  (a_1 + a_2 + 2 a_3) \left(\Gamma_\mu{}^\nu\right)_{\alpha\betabar} +
  (a_1 - a_2) \delta_\mu^\nu \left<\beps_\alpha,
    \bar\beps_\gammabar\right>~,
\end{equation}
which vanishes if and only if $a_1 = a_2 = -a_3$.  In other words, the
only possible (nontrivial) cocycle must be proportional to
\begin{equation}
  \Theta' = P^\mu \wedge Q^\alpha \otimes \Gamma_\mu \cdot Q_\alpha
  + P^\mu \wedge \bar Q^\alphabar \otimes \Gamma_\mu  \cdot \bar
  Q_\alphabar - Q^\alpha \wedge \bar Q^\betabar \otimes
  \Gamma^{\mu\nu}_{\alpha\betabar} L_{\mu\nu}~,
\end{equation}
whose differential $d\Theta'$ has terms of four types: $QQQ\otimes\bar
Q$, $\bar Q \bar Q \bar Q \otimes Q$, $Q \bar Q \bar Q \otimes \bar Q$
and $Q Q \bar Q \otimes Q$.  As we now show, the first two terms
vanish.  It will suffice to see this for the first term, which has the
form
\begin{equation}
  \half Q^\alpha \wedge Q^\beta \wedge Q^\gamma \otimes
  \Gamma^\mu_{\beta\gamma} \Gamma_\mu \cdot Q_\alpha~.
\end{equation}
By the usual polarisation identity, this will vanish if and only if
for all $\varepsilon \in \Delta_+$,
\begin{equation}
  \label{eq:DiracIIA}
  \left<\varepsilon, \Gamma^\mu  \varepsilon\right> \Gamma_\mu
   \varepsilon = 0~,
\end{equation}
which says that Clifford multiplication by the Dirac current of a
chiral spinor, which is null in ten dimensions, annihilates the
spinor.  This is known to be true, as proved, for instance in
\cite[Appendix~A]{EHJGMHom}.  The second term vanishes for precisely
the same reasons, except that $\varepsilon \in \Delta_-$ now.
It remains to investigate the $Q \bar Q \bar Q \otimes \bar Q$
and $Q Q \bar Q \otimes Q$ terms.  It will suffice to analyse the
former term, say, which is given by
\begin{equation}
  \half Q^\alpha \wedge \bar Q^\betabar \wedge \bar Q^\gammabar
  \otimes \bar Q_\alphabar \left(\Gamma^\mu_{\betabar\gammabar}
    \left(\Gamma_\mu\right)^\alphabar{}_\alpha -
    \Gamma^{\mu\nu}_{\alpha\betabar}
    \left(\Gamma_{\mu\nu}\right)^\alphabar{}_\gammabar\right)~.
\end{equation}
Again, using a polarisation identity, this term will vanish if and
only if the following identity holds
\begin{equation}
  \left<\bar\varepsilon, \Gamma^\mu  \bar \varepsilon\right>
  \Gamma_\mu  \varepsilon - \left<\varepsilon, \Gamma^{\mu\nu}
     \bar \varepsilon\right> \Gamma_{\mu\nu}  \bar
  \varepsilon \stackrel{?}{=} 0 \qquad \forall \varepsilon \in
  \Delta_+,~ \bar \varepsilon \in \Delta_-~.
\end{equation}
It is not hard to check that this is not true in general, thus proving
the rigidity of the IIA Poincaré superalgebra.

The rigidity of the IIA Poincaré superalgebra might come as a surprise
due to the existence of massive supergravities
\cite{HLW,Romans-Massive} which deform IIA supergravity by a mass
parameter.  These deformations are such that as the mass parameter
tends to zero one recovers IIA supergravity, whence any background of
such a massive supergravity tends to a IIA supergravity background in
that limit and hence we would expect to find the massive supergravity
background among the deformations of the IIA background, with a
similar situation reflecting itself in their superalgebras.  The fact
that the IIA Poincaré superalgebra is rigid is consistent with the
non-existence of maximally supersymmetric backgrounds for massive
supergravities, which is demonstrated in \cite{FOPMax} for the Romans
theory.  We are not aware of a similar result for the massive IIA
theory of \cite{HLW}, but our rigidity results imply that none exists
which tends in the massless limit to Minkowski spacetime.

\subsection{Rigidity of the D0-brane superalgebra}
\label{sec:D0}

The Killing superalgebra of the D0-brane background is the
subsuperalgebra of the IIA Poincaré superalgebra $\fk = \fk_0 \oplus
\fk_1$, with $\fk_0 = W \oplus \fso(W^\perp)$, with $W$ the
lorentzian line spanned by $\be_0$, and $\fk_1$ isomorphic
to the subspace $\Delta_{\D0} \subset \Delta_+ \oplus \Delta_-$ defined
as the graph of the linear map $\Gamma^0:\Delta_+ \to \Delta_-$.  We
let $\be_a$ span $W^\perp$.  The corresponding basis for $\fk_0$ are
$P$ and $L_{ab}$.  We let $\beps_\alpha$ denote a basis for $\Delta_+$
and $\bpsi_\alpha := \tfrac1{\sqrt{2}} \begin{pmatrix}\beps_\alpha\\
  \Gamma^0  \beps_\alpha\end{pmatrix}$ be a basis for
$\Delta_{\D0}$.  We will let $Q_\alpha$ denote the corresponding basis
for $\fk_1$.  The nonzero Lie brackets are those of $\fso(W^\perp)$
and
\begin{equation}
  \label{eq:IIAD0KSA}
  \begin{aligned}[m]
    [L_{ab},Q_\alpha] &= \half \Gamma_{ab} \cdot Q_\alpha\\
    [Q_\alpha, Q_\beta] &= \Gamma^0_{\alpha\beta} P~,
  \end{aligned}
\end{equation}
where
\begin{equation}
  \Gamma^0_{\alpha\beta} := \langle\bpsi_\alpha, \Gamma^0 
    \bpsi_\beta\rangle = \left<\beps_\alpha, \Gamma^0 
    \beps_\beta\right>~.
\end{equation}

The ideal $I<\fk$ is spanned by $P$ and $Q_\alpha$, whereas the
semisimple factor $\fs = \fso(W^\perp)$ is spanned by the $L_{ab}$.
Letting $P^*$ and $Q^\alpha$ denote the canonical dual basis for
$I^*$, the differential in the deformation complex $C^\bullet =
C^\bullet(I;\fk)^{\fs}$ is determined uniquely by the following
relations
\begin{equation}
  \begin{aligned}[m]
    d P^* &= \half \Gamma^0_{\alpha\beta} Q^\alpha \wedge Q^\beta\\
    d Q^\alpha &= 0\\
    d P &= 0 \\
    d Q_\alpha &= - \Gamma^0_{\alpha\beta} Q^\beta \otimes P\\
    d L_{ab} &= \half Q^\alpha \otimes \Gamma_{ab} \cdot Q_\alpha~.
  \end{aligned}
\end{equation}

The space of $0$-cochains is spanned by $P$, whence $H^0(\fk;\fk)
\cong \RR$.  There is a two-dimensional space of $1$-cochains, spanned
by $P^* \otimes P$ and $Q^\alpha \otimes Q_\alpha$ and a
two-dimensional space of $2$-cochains, spanned by $P^* \wedge Q^\alpha
\otimes Q_\alpha$ and $Q^\alpha \wedge Q^\beta \otimes
\Gamma^0_{\alpha\beta} P$.  Computing the differential $d:C^1 \to
C^2$, we find that $H^1(\fk;\fk) \cong \RR$, with representative
cocycle $2P^* \otimes P - Q^\alpha \otimes Q_\alpha$.  This means that
$\dim B^2 = 1$ and since the differential $d:C^2 \to C^3$ is not
identically zero, one concludes that $H^2(\fk;\fk) = 0$, thus proving
the rigidity of the D0-brane superalgebra.

\subsection{A deformation of the fundamental string superalgebra}
\label{sec:IIAF1}

The Killing superalgebra of the IIA fundamental string solution is the
subsuperalgebra $\fk = \fk_0 \oplus \fk_1$ of the IIA Poincaré
superalgebra associated to a decomposition of the ten-dimensional
lorentzian vector space $V = W \oplus W^\perp$, where $W$ is a
two-dimensional lorentzian subspace corresponding to the string
worldsheet.  This means that $\fk_0 = \fso(W) \oplus W \oplus
\fso(W^\perp)$ and $\fk_1$ is isomorphic to the subspace $\Delta_{\F1}
\subset \Delta_+ \oplus \Delta_-$ defined by
\begin{equation}
  \Delta_{\F1} = \left\{ \begin{pmatrix} \varepsilon_+ \\
      \varepsilon_- \end{pmatrix} \in \Delta_+ \oplus \Delta_- \middle
    | \bnu_W  \varepsilon_\pm = \pm \varepsilon_\pm \right\}~,
\end{equation}
where the volume element $\bnu_W$ obeys $\bnu_W^2 = + \1$ and is
skewsymmetric relative to the spinor inner product.  Alternatively we
may think of $\Delta_{\F1}$ as the subspace of spinors having
positive-chirality under $\fso(W^\perp)$.  We let $\be_\mu$ and
$\be_a$ be a basis for $W$ and $W^\perp$, respectively, and let
$\beps_\alpha$ and $\bar\beps_\alphabar$ span $\Delta_{\F1} \cap
\Delta_\pm$, respectively.  The corresponding basis for $\fk$ is then
$P_\mu$, $L_{\mu\nu} = \epsilon_{\mu\nu} L$, $L_{ab}$, $Q_\alpha$ and
$\bar Q_\alphabar$, with nonzero Lie brackets
\begin{equation}
  \label{eq:IIAF1KSA}
  \begin{aligned}[m]
    [L,Q_\alpha] &= - \half Q_\alpha\\
    [L,\bar Q_\alphabar] &=  \half \bar Q_\alphabar\\
    [L,P_\mu] &=  \epsilon_\mu{}^\nu P_\nu\\
    [Q_\alpha, Q_\beta] &= \Gamma^\mu_{\alpha\beta} P_\mu\\
    [\bar Q_\alphabar, \bar Q_\betabar] &= \Gamma^\mu_{\alphabar\betabar} P_\mu
  \end{aligned}
\end{equation}
in addition to those of $\fso(W^\perp)$, where
\begin{equation}
  \Gamma^\mu_{\alpha\beta} := \left<\beps_\alpha, \Gamma^\mu 
    \beps_\beta\right>
  \qquad\text{and}\qquad
  \Gamma^\mu_{\alphabar\betabar} := \left<\bar\beps_\alphabar, \Gamma^\mu 
    \bar\beps_\betabar\right>~.
\end{equation}

The ideal $I<\fk$ is spanned by $L$, $P_\mu$, $Q_\alpha$ and $\bar
Q_\alphabar$ and the semisimple factor $\fs = \fso(W^\perp)$ by
$L_{ab}$.  The subalgebra $\fr = \fso(W) \oplus \fso(W^\perp)$ spanned
by $L$ and $L_{ab}$ is reductive and the deformation complex
$C^\bullet := C^\bullet(I;\fk)^{\fr}$ consists of $\fr$-invariant
cochains.  Letting $L^*$, $P^\mu$, $Q^\alpha$ and $\bar Q^\alphabar$
denote the canonical dual basis for $I^*$, the differential in the
deformation complex is defined by the following relations:
\begin{equation}
  \begin{aligned}[m]
    d P^\mu &= \half \Gamma^\mu_{\alpha\beta} Q^\alpha \wedge Q^\beta
    + \half \Gamma^\mu_{\alphabar\betabar} \bar Q^\alphabar \wedge
    \bar Q^\betabar + \epsilon^\mu{}_\nu L^* \wedge P^\nu \\
    d Q^\alpha &= \half L^* \wedge Q^\alpha\\
    d \bar Q^\alphabar &= - \half L^* \wedge \bar Q^\alphabar\\
    d L^* &= 0\\
    d P_\mu &= L^* \otimes \epsilon_\mu{}^\nu P_\nu\\
    d Q_\alpha &= - \half L^* \otimes Q_\alpha - \Gamma^\mu_{\alpha\beta} Q^\beta \otimes  P_\mu\\
    d \bar Q_\alphabar &= \half L^* \otimes \bar Q_\alphabar - \Gamma^\mu_{\alphabar\betabar} \bar Q^\betabar \otimes  P_\mu\\
    d L &= - P^\mu \otimes \epsilon_\mu{}^\nu P_\nu - \half Q^\alpha \otimes Q_\alpha + \half \bar Q^\alphabar \otimes \bar Q_\alphabar\\
    d L_{ab} &= \half Q^\alpha \otimes \Gamma_{ab} \cdot Q_\alpha + \half \bar Q^\alphabar \otimes \Gamma_{ab}  \cdot \bar Q_\alphabar~.
  \end{aligned}
\end{equation}

The space of $0$-cochains is $1$-dimensional and spanned by $L$, but
since $dL \neq 0$, $H^0(\fk;\fk) = 0$ and $\dim B^1 = 1$.  The space
$C^1$ is $5$-dimensional and is spanned by $P^\mu \otimes P_\mu$,
$P^\mu \otimes \epsilon_\mu{}^\nu P_\nu$, $L^*\otimes L$, $Q^\alpha
\otimes Q_\alpha$ and $\bar Q^\alphabar \otimes \bar Q_\alphabar$.  As
in the case of the IIB D- and fundamental strings, this can be
understood by the fact that the ideal $I$ is graded by the action of
$2L$ with $L$ having degree $0$, $Q_\alpha$ and $\bar Q_\alphabar$
having degrees $\mp 1$, respectively, and $P_\mu$ having pieces of
degrees $\pm 2$, corresponding to a Witt basis for $W$.  The
$5$-dimensional space of cochains can be thought of as spanned by the
cochains corresponding to the identity maps of each of the five graded
subspaces.  The space $C^2$ of $2$-cochains is $11$-dimensional,
spanned by $L^* \wedge P^\mu \otimes P_\mu$, $L^* \wedge P^\mu \otimes
\epsilon_\mu{}^\nu P_\nu$, $L^* \wedge Q^\alpha \otimes Q_\alpha$,
$L^* \wedge \bar Q^\alphabar \otimes \bar Q_\alphabar$, $P^\mu\wedge
P^\nu\otimes \epsilon_{\mu\nu} L$, $P^\mu \wedge Q^\alpha \otimes
\Gamma_\mu \cdot Q_\alpha$, $P^\mu \wedge \bar Q^\alphabar \otimes
\Gamma_\mu \cdot \bar Q_\alphabar$, $Q^\alpha \wedge Q^\beta \otimes
\Gamma^\mu_{\alpha\beta} P_\mu$, $\bar Q^\alphabar \wedge \bar
Q^\betabar \otimes \Gamma^\mu_{\alphabar\betabar} P_\mu$, $Q^\alpha
\wedge \bar Q^\betabar \otimes (\bnu_W)_{\alpha\betabar} L$ and
$Q^\alpha \wedge \bar Q^\betabar \otimes \Gamma^{ab}_{\alpha\betabar}
L_{ab}$.

Computing the differential $d: C^1 \to C^2$, we find that
$H^1(\fk;\fk) \cong \RR$, with representative cocycle $\varphi:= 2
P^\mu \otimes P_\mu - Q^\alpha \otimes Q_\alpha - \bar Q^\alphabar
\otimes \bar Q_\alphabar$, whence $\dim B^2 = 3$, spanned by $d
(Q^\alpha \otimes Q_\alpha)$, $d( \bar Q^\alphabar\otimes \bar
Q_\alphabar)$ and $ d(L^* \otimes L)$, say.  Similarly, computing
$d:C^2 \to C^3$ we find that $H^2(\fk;\fk)\cong \RR$, with
representative cocycle $L^* \wedge \varphi$.  This infinitesimal
deformation integrates to a one-parameter family of Lie superalgebras
with brackets
\begin{equation}
  \label{eq:IIAF1deformed}
  \begin{aligned}[m]
    [L,Q_\alpha] &= (t - \half) Q_\alpha \\
    [L,\bar Q_\alphabar] &= (t + \half) \bar Q_\alphabar \\
    [L,P_\mu] &=  2 t P_\mu + \epsilon_\mu{}^\nu P_\nu\\
    [Q_\alpha, Q_\beta] &= \Gamma^\mu_{\alpha\beta} P_\mu\\
    [\bar Q_\alphabar, \bar Q_\betabar] &=
    \Gamma^\mu_{\alphabar\betabar} P_\mu~,
  \end{aligned}
\end{equation}
in addition to those involving $\fso(W^\perp)$ which remain
undeformed.  This deformation consists of changing the $L$-weight of
the generators in the Lie superalgebra in such a way that the $QQ$ and
$\bar Q\bar Q$ brackets remains invariant, just as in the IIB
fundamental string.  As in the cases of the IIB D- and fundamental
strings, this deformation is reminiscent of the construction of
two-dimensional topological conformal field theories via twisting.

The fundamental string solution arises as the Kaluza--Klein reduction
of the M2 brane along a translational symmetry of the brane
worldvolume.  In view of this, one might expect that the deformation
of the M2 superalgebra found in \cite{JMFSuperDeform} might induce a
deformation of the fundamental string superalgebra, yet the
deformation found above in \eqref{eq:IIAF1deformed} is not the
reduction of the one for the M2 brane superalgebra.  If we take the
geometric origin of the deformed M2 superalgebra at face value, the
worldvolume of the putative deformed M2 brane is now $\AdS_3$ and it
follows from the results of \cite{FigSimAdS} that no quotient (regular
or singular) of $\AdS_3$ preserves all the supersymmetries, whence the
superalgebra of the Kaluza--Klein reduction of such a deformed
M2-brane would be of strictly smaller dimension than that of the
fundamental string superalgebra and hence would not appear among its
deformations.

\subsection{A deformation of the D2-brane superalgebra}
\label{sec:D2}

The D2-brane Killing superalgebra $\fk$ is the subsuperalgebra of the
IIA Poincaré superalgebra corresponding to the split $V = W \oplus
W^\perp$, with $W$ a $3$-dimensional lorentzian subspace.  The even
subalgebra $\fk_0 = \fso(W) \oplus W \oplus \fso(W^\perp)$ and the odd
subspace $\fk_1$ is isomorphic to the subspace $\Delta_{\D2} \subset
\Delta_+ \oplus \Delta_-$ defined as the graph of $\bnu_W : \Delta_+
\to \Delta_-$.  The linear map $\bnu_W$ is symmetric relative to the
spinor inner product and obeys $\bnu_W^2 = +\1$.  Let $\be_\mu$ and
$\be_a$ span $W$ and $W^\perp$, respectively.  Let $\beps_\alpha$ span
$\Delta_+$ so that $\bpsi_\alpha :=
\tfrac1{\sqrt{2}} \begin{pmatrix}\beps_\alpha\\ \bnu_W 
  \beps_\alpha\end{pmatrix}$ span $\Delta_{\D2}$.  The corresponding
basis for $\fk$ is $P_\mu$, $L_{\mu\nu}$, $L_{ab}$ and $Q_\alpha$,
with nonzero Lie brackets given, in addition to those of $\fk_0$, by
\begin{equation}
  \label{eq:IIAD2KSA}
  \begin{aligned}[m]
    [L_{\mu\nu},Q_\alpha] &= \half \Gamma_{\mu\nu} \cdot Q_\alpha\\
    [L_{ab},Q_\alpha] &= \half \Gamma_{ab} \cdot Q_\alpha\\
    [Q_\alpha, Q_\beta] &= \Gamma^\mu_{\alpha\beta} P_\mu~,
  \end{aligned}
\end{equation}
where
\begin{equation}
  \Gamma^\mu_{\alpha\beta} := \langle\bpsi_\alpha, \Gamma^\mu 
    \bpsi_\beta\rangle = \left<\beps_\alpha, \Gamma^\mu 
    \beps_\beta\right>~.
\end{equation}
We observe that
\begin{equation}
  \langle\bpsi_\alpha, \Gamma^{\mu\nu}  \bpsi_\beta\rangle  = 
  \left<\beps_\alpha, \Gamma^{\mu\nu} \bnu_W 
    \beps_\beta\right> =: \left(\Gamma^{\mu\nu}
  \bnu_W\right)_{\alpha\beta}
\end{equation}
and, similarly,
\begin{equation}
  \langle\bpsi_\alpha, \Gamma^{ab}  \bpsi_\beta\rangle  = 
  \langle\beps_\alpha, \Gamma^{ab} \bnu_W  \beps_\beta\rangle
  =: (\Gamma^{ab} \bnu_W)_{\alpha\beta} ~,
\end{equation}
whereas
\begin{equation}
    \langle\bpsi_\alpha, \Gamma^a  \bpsi_\beta\rangle = 0 =
    \langle\bpsi_\alpha, \Gamma^{\mu a}  \bpsi_\beta\rangle~.
\end{equation}

The ideal $I<\fk$ is spanned by $P_\mu$ and $Q_\alpha$, whereas the
semisimple factor $\fs = \fso(W) \oplus \fso(W^\perp)$ is spanned by
$L_{\mu\nu}$ and $L_{ab}$.  The canonical dual basis for $I^*$ is
$P^\mu$ and $Q^\alpha$, relative to which the differential in the
deformation complex $C^\bullet := C^\bullet(I;\fk)^{\fs}$ is defined
by the following relations
\begin{equation}
  \begin{aligned}[m]
    d P^\mu &= \half \Gamma^\mu_{\alpha\beta} Q^\alpha \wedge Q^\beta\\
    d Q^\alpha &= 0\\
    d P_\mu &= 0\\
    d Q_\alpha &= - \Gamma^\mu_{\alpha\beta} Q^\beta \otimes  P_\mu\\
    d L_{\mu\nu} &=  \eta_{\mu\rho} P^\rho \otimes P_\nu -
    \eta_{\nu\rho} P^\rho \otimes P_\mu + \half Q^\alpha \otimes
    \Gamma_{\mu\nu} \cdot Q_\alpha\\
    d L_{ab} &= \half Q^\alpha \otimes \Gamma_{ab} \cdot Q_\alpha~.
  \end{aligned}
\end{equation}

There are no $0$-cochains since $\fk^{\fs} = 0$.  The space of
$1$-cochains is spanned by $P^\mu \otimes P_\mu$, $Q^\alpha \otimes
Q_\alpha$ and $P^\rho \otimes \epsilon_\rho{}^{\mu\nu} L_{\mu\nu}$,
corresponding to the identity maps $W \to W$, $\Delta_{\D2} \to
\Delta_{\D2}$ and the composition $W \to \Lambda^2 W \cong \fso(W)$
where the first map is induced by the Hodge star.  The space of
$2$-cochains is $6$-dimensional and is spanned by $P^\mu\wedge
P^\nu\otimes L_{\mu\nu}$, $P^\mu\wedge P^\nu\otimes
\epsilon_{\mu\nu}{}^\rho P_\rho$, $P^\mu \wedge Q^\alpha \otimes
\Gamma_\mu \cdot Q_\alpha$, $Q^\alpha \wedge Q^\beta \otimes
\Gamma^\mu_{\alpha\beta} P_\mu$, $Q^\alpha \wedge Q^\beta \otimes
\left(\Gamma^{\mu\nu} \bnu_W\right)_{\alpha\beta} L_{\mu\nu}$ and
$Q^\alpha \wedge Q^\beta \otimes \left(\Gamma^{ab}
  \bnu_W\right)_{\alpha\beta} L_{ab}$.  Notice that the cochain
$P^\rho \wedge Q^\alpha \otimes \epsilon_\rho{}^{\mu\nu}
\Gamma_{\mu\nu} \cdot Q_\alpha$ is already contained in the above
span, since
\begin{equation}
  \label{eq:hodge3}
  \epsilon_{\mu\nu\rho} \Gamma^\rho \cdot Q_\alpha = - \Gamma_{\mu\nu}
  \cdot Q_\alpha
  \qquad\text{and}\qquad
  \half \epsilon_{\rho\mu\nu} \Gamma^{\mu\nu} \cdot Q_\alpha =
  \Gamma_\rho \cdot Q_\alpha~.
\end{equation}

Computing the differential $d:C^1 \to C^2$ we find that $H^1(\fk;\fk)
\cong \RR$, with representative cocycle $2P^\mu \otimes P_\mu -
Q^\alpha \otimes Q_\alpha$.  This means that $\dim B^2 = 2$, spanned
by $Q^\alpha \wedge Q^\beta \otimes \Gamma^\mu_{\alpha\beta}
P_\mu$ and
\begin{multline}
  \label{eq:coboundary}
  d \left( P^\rho \otimes \epsilon_\rho{}^{\mu\nu} L_{\mu\nu} \right)
  = - \half \left(\Gamma^{\mu\nu} \bnu_W\right)_{\alpha\beta}
  Q^\alpha \wedge Q^\beta \otimes L_{\mu\nu}\\
  - 2 \epsilon_{\mu\nu}{}^\rho P^\mu \wedge P^\nu \otimes P_\rho -
  P^\mu \wedge Q^\alpha \otimes \Gamma_\mu \cdot Q_\alpha~.
\end{multline}

Computing the differential $d:C^2 \to C^3$ we find, in addition to the
coboundaries,
\begin{equation}
  \begin{aligned}[m]
    d \left(P^\mu\wedge P^\nu\otimes L_{\mu\nu} \right) & =
    \Gamma^\mu_{\alpha\beta} Q^\alpha \wedge Q^\beta \wedge P^\nu
    \otimes L_{\mu\nu} + \half P^\mu \wedge P^\nu \wedge Q^\alpha
    \otimes \Gamma_{\mu\nu} \cdot Q_\alpha\\
    d \left(P^\mu\wedge P^\nu\otimes \epsilon_{\mu\nu}{}^\rho P_\rho
    \right) &= - \left(\Gamma_\mu{}^\nu 
      \bnu_W\right)_{\alpha\beta} Q^\alpha \wedge Q^\beta \wedge P^\mu
    \otimes P_\nu\\
    d \left(P^\mu \wedge Q^\alpha \otimes \Gamma_\mu \cdot
      Q_\alpha\right) &=  \half \Gamma^\mu_{\alpha\beta} Q^\alpha 
    \wedge Q^\beta \wedge Q^\gamma \otimes \Gamma_\mu \cdot Q_\gamma\\
    & \quad {} + \left(\Gamma_\mu{}^\nu 
      \bnu_W\right)_{\alpha\beta} Q^\alpha \wedge Q^\beta \wedge P^\mu
    \otimes P_\nu\\
    d \left(Q^\alpha \wedge Q^\beta \otimes \left(\Gamma^{\mu\nu}
        \bnu_W\right)_{\alpha\beta} L_{\mu\nu}\right) &= 2 \left(\Gamma_\mu{}^\nu 
      \bnu_W\right)_{\alpha\beta} Q^\alpha \wedge Q^\beta \wedge P^\mu
    \otimes P_\nu\\
    & \quad {} + \half \left(\Gamma^{\mu\nu}
      \bnu_W\right)_{\alpha\beta} Q^\alpha
    \wedge Q^\beta \wedge Q^\gamma \otimes \Gamma_{\mu\nu} \cdot Q_\gamma\\
    d \left(Q^\alpha \wedge Q^\beta \otimes (\Gamma^{ab}
        \bnu_W)_{\alpha\beta} L_{ab}\right) &= \half (\Gamma^{ab}
      \bnu_W)_{\alpha\beta} Q^\alpha \wedge Q^\beta \wedge
    Q^\gamma \otimes \Gamma_{ab} \cdot Q_\gamma~.
  \end{aligned}
\end{equation}
The first term is the only one depending on $L_{\mu\nu}$, whence any
cocycle must be a linear combination of the other cochains
\begin{multline}
  \Theta = a_1 P^\mu\wedge P^\nu\otimes \epsilon_{\mu\nu}{}^\rho
  P_\rho + a_2 P^\mu \wedge Q^\alpha \otimes \Gamma_\mu \cdot Q_\alpha\\
  + a_3 Q^\alpha \wedge Q^\beta \otimes \left(\Gamma^{\mu\nu}
    \bnu_W\right)_{\alpha\beta} L_{\mu\nu} + a_4 Q^\alpha \wedge
  Q^\beta \otimes (\Gamma^{ab} \bnu_W)_{\alpha\beta} L_{ab}~.
\end{multline}
Computing its differential, we find two types of terms: a $PQQ\otimes
P$ term proportional to 
\begin{equation}
  \left(\Gamma_\mu{}^\nu 
      \bnu_W\right)_{\alpha\beta} Q^\alpha \wedge Q^\beta \wedge P^\mu
    \otimes P_\nu \left( -a_1 + a_2 + 2 a_3\right)~,
\end{equation}
and a $QQQ\otimes Q$ term proportional to
\begin{equation}
  Q^\alpha \wedge Q^\beta \wedge Q^\gamma \otimes Q_\delta \left(
    a_2 \Gamma^\mu_{\alpha\beta} \left(\Gamma_\mu
      \bnu_W\right)^\delta{}_\gamma + a_3
    \left(\Gamma^{\mu\nu} \bnu_W\right)_{\alpha\beta}
    \left(\Gamma_{\mu\nu}\right)^\delta{}_\gamma + a_4 
    (\Gamma^{ab} \bnu_W)_{\alpha\beta}
    \left(\Gamma_{ab}\right)^\delta{}_\gamma\right)~.
\end{equation}
The vanishing of the first of the above terms forces $a_1 = a_2 +
2a_3$, whereas the vanishing of the second term is equivalent to
\begin{equation}
  a_2 \left<\varepsilon, \Gamma^\mu  \varepsilon\right> \Gamma_\mu
   \bnu_W  \varepsilon + a_3 \left<\varepsilon,
    \Gamma^{\mu\nu}  \bnu_W  \varepsilon\right>
  \Gamma_{\mu\nu}  \varepsilon + a_4 \langle\varepsilon,
    \Gamma^{ab}  \bnu_W  \varepsilon\rangle
  \Gamma_{ab}  \varepsilon = 0
\end{equation}
for all $\varepsilon \in \Delta_+$.  Using equation
\eqref{eq:fierz10dbrane3}, we can rewrite this as
\begin{equation}
  (a_2 - 8 a_4) \left<\varepsilon, \Gamma^\mu  \varepsilon\right> \Gamma_\mu
   \bnu_W  \varepsilon + (a_3 -a_4) \left<\varepsilon,
    \Gamma^{\mu\nu}  \bnu_W  \varepsilon\right>
  \Gamma_{\mu\nu}  \varepsilon = 0~,
\end{equation}
whereas using \eqref{eq:hodge3} we can finally rewrite this as
\begin{equation}
  (a_2 - 2 a_3 - 6 a_4) \left<\varepsilon, \Gamma^\mu 
    \varepsilon\right> \Gamma_\mu  \bnu_W  \varepsilon =
  0~,
\end{equation}
which vanishes if and only if $a_2 = 2 a_3 + 6 a_4$.  Therefore we see
that there is a two-dimensional space of such cocycles, labelled by
$a_3$ and $a_4$.  The line $a_4 = 0$ is spanned by the coboundary in
equation \eqref{eq:coboundary}, whence we conclude that $H^2(\fk;\fk)
\cong \RR$, with representative cocycle
\begin{equation}
  \label{eq:cocycle}
  6 P^\mu\wedge P^\nu\otimes \epsilon_{\mu\nu}{}^\rho
  P_\rho + 6 P^\mu \wedge Q^\alpha \otimes \Gamma_\mu \cdot Q_\alpha +
  Q^\alpha \wedge Q^\beta \otimes
  (\Gamma^{ab}\bnu_W)_{\alpha\beta} L_{ab}~,
\end{equation}
which spans the line $a_3=0$.  This infinitesimal deformation
integrates to a one-parameter family of Lie superalgebras which, in
addition to the undeformed brackets involving $\fs$, has the following
nonzero brackets:
\begin{equation}
  \begin{aligned}[m]
    [P_\mu,P_\nu] &= 12 t \epsilon_{\mu\nu}{}^\rho P_\rho\\
    [P_\mu,Q_\alpha] &= -6t \Gamma_\mu \cdot Q_\alpha\\
    [Q_\alpha, Q_\beta] &= \Gamma^\mu_{\alpha\beta} P_\mu - 2t
    (\Gamma^{ab}\bnu_W)_{\alpha\beta} L_{ab}~.
  \end{aligned}
\end{equation}
It is not hard to show that the Jacobi identity is satisfied: only the
$O(t^2)$ terms need to be checked since, by construction, the others
are satisfied.  The Jacobi identity breaks up into several types of
term: $PPP$, $PPQ$, $PQQ$ and $QQQ$, the first three of which vanish
trivially and the last vanishes by virtue of the Fierz identity
\eqref{eq:fierz10dbrane3}.  For $t\neq 0$, we may rescale the
generators $P$ and $Q$ in order to bring the above Lie superalgebra to
the following form
\begin{equation}
  \label{eq:IIAD2deformed}
  \begin{aligned}[m]
    [P_\mu,P_\nu] &= -2 \epsilon_{\mu\nu}{}^\rho P_\rho\\
    [P_\mu,Q_\alpha] &= \Gamma_\mu \cdot Q_\alpha\\
    [Q_\alpha, Q_\beta] &= \pm \left( \Gamma^\mu_{\alpha\beta} P_\mu +
      \tfrac13 (\Gamma^{ab}\bnu_W)_{\alpha\beta} L_{ab}
    \right)~.
  \end{aligned}
\end{equation}
As in the case of the M2-brane superalgebra deformation in
\cite{JMFSuperDeform}, the choice of sign corresponds to a duality
relating two real forms of the same complex superalgebra,
corresponding to multiplying the odd generators by $i$.  The even
subalgebra $\fk_0 \cong \fso(2,2) \oplus \fso(7)$ with $\fk_1$
transforming as $\Delta^{(2,2)}_+ \otimes \Delta^{(7)}$, with
$\Delta^{(2,2)}_+$ the real $2$-dimensional positive-chirality spinor
representation of $\fso(2,2)$ and $\Delta^{(7)}$ the real
$8$-dimensional spinor representation of $\fso(7)$.  Let us change
basis from $L_{\mu\nu}$ to $L'_{\mu\nu} := L_{\mu\nu} + \half
\epsilon_{\mu\nu}{}^\rho P_\rho$.  We notice that $L'_{\mu\nu}$ commute
with the supercharges and hence with $P_\mu$ and that they span an 
$\fso(2,1)$ subalgebra.  The remaining generators $P_\mu$, $L_{ab}$
and $Q_\alpha$ span a simple subsuperalgebra of $\fk$ isomorphic to
the exceptional Lie \emph{super}algebra $\ff(4)$
\cite{KacSuperSketch}.  Therefore, as an abstract Lie superalgebra,
the deformed Lie superalgebra in \eqref{eq:IIAD2deformed} is
isomorphic to $\fso(2,1) \oplus \ff(4)$.

\subsubsection{Deformations of the delocalised M2-brane superalgebra}
\label{sec:DM2}

The D2-brane solution of type IIA supergravity arises via
Kaluza--Klein reduction from an M2-brane which has been delocalised
along one transverse direction.  In this section we will show that the
deformation of the D2-brane superalgebra \eqref{eq:IIAD2deformed}
found above has its origin in a deformation of the delocalised M2-brane
superalgebra.  As Kaluza--Klein reduction \emph{is} a geometric
procedure, this results lends support to the hypothesis that these
deformations have a geometric origin.

Let $V$ be here an eleven-dimensional lorentzian vector space and we
decompose it as $V = W \oplus U \oplus \RR\be_\ten$, where $W$ is a
three-dimensional lorentzian subspace, corresponding to the membrane
worldvolume, $W^\perp = U \oplus \RR\be_\ten$ is the transverse space,
where $\be_\ten$ denotes the delocalised eleventh direction.
Delocalisation means that the metric and four-form do not depend on
the eleventh coordinate.  The symmetric delocalised M2-brane has
superalgebra $\fk = \fk_0 \oplus \fk_1$, where $\fk_0 = \fso(W) \oplus
W \oplus \fso(U) \oplus \RR$, with the $\RR$ subalgebra corresponds to
translations along the eleventh direction, and where $\fk_1$ is
isomorphic to the subspace $\Delta_{\DM2} \subset \Delta$ of the
spinor representation consisting of spinors $\varepsilon \in \Delta$
obeying $\bnu_W \varepsilon = \varepsilon$, where $\bnu_W$ is the
volume element associated to $W$.  It is convenient, in order to
compare with the IIA results, to break up $\Delta = \Delta_+ \oplus
\Delta_-$ into eigenspaces of $\Gamma_\ten$, which is in fact the same
split as the one induced by chirality in ten dimensions. The subspace
$\Delta_{\DM2}$ defined above is then the graph of $\bnu_W : \Delta_+
\to \Delta_-$, so that if $\beps_\alpha$ is a basis for $\Delta_+$,
then $\bpsi_\alpha = \frac{1}{\sqrt{2}}\begin{pmatrix}\beps_\alpha \\
  \bnu_W \beps_\alpha\end{pmatrix}$ is a basis for $\Delta_{\DM2}$.
Let $P_\mu$, $P=P_\ten$, $L_{\mu\nu}$, $L_{ab}$ and $Q_\alpha$ denote
a basis for the $\fk$, whose nonzero Lie brackets are given, in
addition to those of $\fk_0$, by
\begin{equation}
  \label{eq:DM2KSA}
  \begin{aligned}[m]
    [L_{\mu\nu},Q_\alpha] &= \half \Gamma_{\mu\nu} \cdot Q_\alpha\\
    [L_{ab},Q_\alpha] &= \half \Gamma_{ab} \cdot Q_\alpha\\
    [Q_\alpha, Q_\beta] &= \Gamma^\mu_{\alpha\beta} P_\mu~,
  \end{aligned}
\end{equation}
where
\begin{equation}
  \Gamma^\mu_{\alpha\beta} := \langle\bpsi_\alpha, \Gamma^\mu 
    \bpsi_\beta\rangle = \left<\beps_\alpha, \Gamma^\mu
    \beps_\beta\right>~.
\end{equation}
We observe that
\begin{equation}
  \langle\bpsi_\alpha, \Gamma^{\mu\nu}  \bpsi_\beta\rangle  = 
  \left<\beps_\alpha, \Gamma^{\mu\nu} \bnu_W 
    \beps_\beta\right> =: \left(\Gamma^{\mu\nu}
  \bnu_W\right)_{\alpha\beta}
\end{equation}
and, similarly,
\begin{equation}
  \langle\bpsi_\alpha, \Gamma^{ab}  \bpsi_\beta\rangle  = 
  \langle\beps_\alpha, \Gamma^{ab} \bnu_W  \beps_\beta\rangle
  =: (\Gamma^{ab} \bnu_W)_{\alpha\beta}~.
\end{equation}

The ideal $I<\fk$ is spanned by $P$, $P_\mu$ and $Q_\alpha$, whereas
the semisimple factor $\fs = \fso(W) \oplus \fso(U)$ is spanned by
$L_{\mu\nu}$ and $L_{ab}$.  The canonical dual basis for $I^*$ is
$P^*$, $P^\mu$ and $Q^\alpha$, relative to which the differential in the
deformation complex $C^\bullet := C^\bullet(I;\fk)^{\fs}$ is defined
by the following relations
\begin{equation}
  \begin{aligned}[m]
    d P^* &= 0\\
    d P^\mu &= \half \Gamma^\mu_{\alpha\beta} Q^\alpha \wedge Q^\beta\\
    d Q^\alpha &= 0\\
    d P &= 0\\
    d P_\mu &= 0\\
    d Q_\alpha &= - \Gamma^\mu_{\alpha\beta} Q^\beta \otimes  P_\mu\\
    d L_{\mu\nu} &=  \eta_{\mu\rho} P^\rho \otimes P_\nu -
    \eta_{\nu\rho} P^\rho \otimes P_\mu + \half Q^\alpha \otimes
    \Gamma_{\mu\nu} \cdot Q_\alpha\\
    d L_{ab} &= \half Q^\alpha \otimes \Gamma_{ab} \cdot Q_\alpha~.
  \end{aligned}
\end{equation}

The space of $0$-cochains is spanned by $P$, which is a cocycle,
whence $H^0(\fk;\fk) \cong \RR$.  The space of $1$-cochains is
$4$-dimensional, spanned by $P^* \otimes P$, $P^\mu \otimes P_\mu$,
$Q^\alpha \otimes Q_\alpha$ and $P^\rho \otimes
\epsilon_\rho{}^{\mu\nu} L_{\mu\nu}$.  The space of $2$-cochains is
$9$-dimensional, spanned by $P^* \wedge P^\mu \otimes P_\mu$,
$P^* \wedge Q^\alpha \otimes Q_\alpha$, $P^* \wedge P^\rho \otimes
\epsilon_\rho{}^{\mu\nu} L_{\mu\nu}$, $P^\mu\wedge P^\nu\otimes
L_{\mu\nu}$, $P^\mu\wedge P^\nu\otimes \epsilon_{\mu\nu}{}^\rho
P_\rho$, $P^\mu \wedge Q^\alpha \otimes \Gamma_\mu \cdot Q_\alpha$,
$Q^\alpha \wedge Q^\beta \otimes \Gamma^\mu_{\alpha\beta} P_\mu$,
$Q^\alpha \wedge Q^\beta \otimes \left(\Gamma^{\mu\nu}
  \bnu_W\right)_{\alpha\beta} L_{\mu\nu}$ and $Q^\alpha \wedge Q^\beta
\otimes \left(\Gamma^{ab} \bnu_W\right)_{\alpha\beta} L_{ab}$.  Notice
that the cochain $P^\rho \wedge Q^\alpha \otimes \epsilon_\rho{}^{\mu\nu}
\Gamma_{\mu\nu} \cdot Q_\alpha$ is already contained in the above
span by equation \eqref{eq:hodge3}.

Computing the differential $d:C^1 \to C^2$, we see that $H^1(\fk;\fk)
\cong \RR^2$, with representative cocycles $P^* \otimes P$ and
$\varphi:= 2 P^\mu \otimes P_\mu- Q^\alpha \otimes Q_\alpha$.  This
means that $\dim B^2 = 2$, spanned by $d (Q^\alpha \otimes Q_\alpha)$,
say, and $d(P^\rho \otimes \epsilon_\rho{}^{\mu\nu} L_{\mu\nu})$ which
is given formally by the expression in \eqref{eq:coboundary}, suitably
reinterpreted for the present situation.  Computing the differential
$d:C^2 \to C^3$ we find that $H^2(\fk;\fk) \cong \RR^2$, with
representative cocycles $P^* \wedge \varphi$ and the one formally
given by the expression \eqref{eq:cocycle}, again suitable
reinterpreted.  Only two lines in $H^2(\fk;\fk)$ give rise to
integrable deformations: the ones spanned by the two cocycles listed
above.  The first one gives rise to the one-parameter family of Lie
algebras where $P$ is no longer central but instead is a grading
element with brackets
\begin{equation}
  \label{eq:DM2deformed1}
  \begin{aligned}[m]
    [P,Q_\alpha] &= t Q_\alpha\\
    [P,P_\mu] &= 2t P_\mu~,
  \end{aligned}
\end{equation}
which for $t\neq 0$ can always be rescaled to set $t=1$.  The second
one is more interesting, since it is the one which reduces to the
deformation \eqref{eq:IIAD2deformed} found above for the D2-brane
superalgebra.  The cocycle is formally identical and so are the
brackets, which can be read off from \eqref{eq:IIAD2deformed}.

\subsection{Rigidity of the D4-brane superalgebra}
\label{sec:D4}

The D4-brane Killing superalgebra $\fk = \fk_0 \oplus \fk_1$ is the
subsuperalgebra of the IIA Poincaré superalgebra corresponding to the
split $V = W \oplus W^\perp$, with $W$ five-dimensional lorentzian.
This means that $\fk_0 = \fso(W) \oplus W \oplus \fso(W^\perp)$ and
$\fk_1$ is isomorphic to the subspace $\Delta_{\D4} \subset \Delta_+
\oplus \Delta_-$ defined as the graph of the linear map $\bnu_W :
\Delta_+ \to \Delta_-$ defined by the Clifford action of the volume
element of $W$, which is skewsymmetric under the spinor inner product
and obeys $\bnu_W^2 = - \1$.

We let $\be_\mu$ and $\be_a$ span $W$ and $W^\perp$, respectively and
let $P_\mu$, $L_{\mu\nu}$ and $L_{ab}$ denote the corresponding basis
for $\fk_0$.  Let $\beps_\alpha$ span $\Delta_+$
and let $\bpsi_\alpha := \tfrac1{\sqrt{2}} \begin{pmatrix}\beps_\alpha\\
  \bnu_W  \beps_\alpha\end{pmatrix}$ be a basis for
$\Delta_{\D4}$.  We will let $Q_\alpha$ denote the corresponding basis
for $\fk_1$.  The nonzero Lie brackets are those of $\fso(W) \oplus
\fso(W^\perp)$ and in addition the following:
\begin{equation}
  \label{eq:IIAD4KSA}
  \begin{aligned}[m]
    [Q_\alpha,Q_\beta] &= \Gamma^\mu_{\alpha\beta} P_\mu\\
    [L_{\mu\nu}, Q_\alpha] &= \half \Gamma_{\mu\nu} \cdot Q_\alpha\\
    [L_{ab}, Q_\alpha] &= \half \Gamma_{ab} \cdot Q_\alpha~,
  \end{aligned}
\end{equation}
where, as usual,
\begin{equation}
  \Gamma^\mu_{\alpha\beta} := \langle\bpsi_\alpha, \Gamma^\mu 
    \bpsi_\beta\rangle =  \left<\beps_\alpha, \Gamma^\mu 
    \beps_\beta\right>~.
\end{equation}

The ideal $I<\fk$ is spanned by $P_\mu$ and $Q_\alpha$ and the
semisimple factor $\fs$ by $L_{\mu\nu}$ and $L_{ab}$.  Let $P^\mu$ and
$Q^\alpha$ be the canonical dual basis for $I^*$.  Relative to these,
the differential in the deformation complex $C^\bullet :=
C^\bullet(I;\fk)^{\fs}$ is defined uniquely by the following
relations:
\begin{equation}
  \begin{aligned}[m]
    d P^\mu &= \Gamma^\mu_{\alpha\beta} Q^\alpha \wedge Q^\beta\\
    d Q^\alpha &= 0\\
    d P_\mu &= 0 \\
    d Q_\alpha &= - \Gamma^\mu_{\alpha\beta} Q^\beta \otimes P_\mu\\
    d L_{\mu\nu} &=  \eta_{\mu\rho} P^\rho \otimes P_\nu -
    \eta_{\nu\rho} P^\rho \otimes P_\mu + \half Q^\alpha \otimes
    \Gamma_{\mu\nu} \cdot Q_\alpha\\
    d L_{ab} &= \half Q^\alpha \otimes \Gamma_{ab} \cdot Q_\alpha~.
  \end{aligned}
\end{equation}

There are no $0$-cochains, whereas the space of $1$-cochains is
$2$-dimensional spanned by the cochains corresponding to the identity
maps $W \to W$ and $\Delta_{\D4} \to \Delta_{\D4}$, namely $P^\mu
\otimes P_\mu$ and $Q^\alpha \otimes Q_\alpha$.  The space of
$2$-cochains is now $3$-dimensional, spanned by $P^\mu \wedge P^\nu
\otimes L_{\mu\nu}$, $P^\mu \wedge Q^\alpha \otimes \Gamma_\mu  \cdot
Q_\alpha$ and $Q^\alpha \wedge Q^\beta \otimes
\Gamma^\mu_{\alpha\beta} P_\mu$.  Computing the differential $d:C^1
\to C^2$ we find that $H^1(\fk;\fk) \cong \RR$, with representative
cocycle $2P^* \otimes P - Q^\alpha \otimes Q_\alpha$.  This means that
$\dim B^2 = 1$, spanned by $d (Q^\alpha\otimes Q_\alpha)$, say.  The
only possible $2$-cocycle is a linear combination
\begin{equation}
  \Theta = a_1 P^\mu \wedge P^\nu \otimes L_{\mu\nu} + a_2 P^\mu
  \wedge Q^\alpha \otimes \Gamma_\mu \cdot Q_\alpha~.
\end{equation}
The first term in the right-hand side is the only one whose
differential contains $L_{\mu\nu}$, whence $d\Theta=0$ forces $a_1 =
0$.  Computing the differential of the second term, we find
\begin{equation}
  d \left( P^\mu \wedge Q^\alpha \otimes \Gamma_\mu \cdot Q_\alpha
  \right) = \half \Gamma^\mu_{\alpha\beta} Q^\alpha \wedge Q^\beta
  \wedge Q^\gamma \otimes \Gamma_\mu \cdot Q_\gamma - P^\mu \wedge
  Q^\alpha \wedge Q^\gamma \otimes \left(\Gamma_\mu 
    \bnu_W\right)^\beta{}_\alpha \Gamma^\nu_{\beta\gamma} P_\nu~.
\end{equation}
The first term vanishes because of identity \eqref{eq:DiracIIA} and
the fact that $\left<\varepsilon, \Gamma^a \varepsilon\right> = 0$,
but the vanishing of the second term requires
\begin{equation}
  \left(\Gamma_\mu  \bnu_W\right)^\beta{}_\alpha
  \Gamma^\nu_{\beta\gamma} + (\alpha\leftrightarrow\gamma)
  \stackrel{?}{=} 0~,
\end{equation}
which by the usual polarisation identity, is equivalent to
\begin{equation}
  \left<\Gamma_\mu  \bnu_W  \varepsilon, \Gamma^\nu
    \varepsilon\right> \stackrel{?}{=} 0 \qquad \forall \varepsilon
  \in \Delta_+~,
\end{equation}
or, equivalently,
\begin{equation}
  \left<\varepsilon, \Gamma_\mu  \Gamma^\nu  \bnu_W 
    \varepsilon\right> \stackrel{?}{=} 0 \qquad \forall \varepsilon
  \in \Delta_+~.
\end{equation}
Using the Clifford algebra and the fact that $\left<\varepsilon,
  \Gamma_{\mu\nu} \bnu_W \varepsilon\right> =0$ for all
$\varepsilon\in\Delta_+$, we are left with
\begin{equation}
  \left<\varepsilon, \bnu_W  \varepsilon\right> \stackrel{?}{=} 0
  \qquad \forall \varepsilon \in \Delta_+~,
\end{equation}
which is patently false, thus proving the rigidity of the D4-brane
superalgebra.

\subsection{Rigidity of the NS 5-brane superalgebra}
\label{sec:IIANS5}

The IIA NS5-brane Killing superalgebra $\fk = \fk_0 \oplus \fk_1$ is
the subsuperalgebra of the IIA Poincaré superalgebra corresponding to
a decomposition $V = W \oplus W^\perp$, where $W$ is a six-dimensional
lorentzian subspace corresponding to the brane worldvolume.  In other
words, $\fk_0 = \fso(W) \oplus W \oplus \fso(W^\perp)$ and $\fk_1$ is
isomorphic to the subspace $\Delta_{\NS5} \subset \Delta_+ \oplus
\Delta_-$ defined as the $+1$ eigenspace of the Clifford action of the
volume element $\bnu_W$ of $W$, which obeys $\bnu_W^2 = +\1$ and is
skewsymmetric relative to the spinor (symplectic) inner product.

Let $\be_\mu$ and $\be_a$ span $W$ and $W^\perp$, respectively and let
$\beps_\alpha$ and $\bar\beps_\alphabar$ denote bases for
$\Delta_{\NS5} \cap \Delta_\pm$, respectively.  The corresponding
basis for $\fk$ is $P_\mu$, $L_{\mu\nu}$, $L_{ab}$, $Q_\alpha$ and
$\bar Q_\alphabar$, and the Lie brackets are, in addition to those of
$\fk_0$, the following:
\begin{equation}
  \label{eq:IIANS5KSA}
  \begin{aligned}[m]
    [L_{\mu\nu},Q_\alpha] &= \half \Gamma_{\mu\nu} \cdot Q_\alpha\\
    [L_{\mu\nu},\bar Q_\alphabar] &= \half \Gamma_{\mu\nu}  \cdot \bar Q_\alphabar\\
    [L_{ab},Q_\alpha] &= \half \Gamma_{ab} \cdot Q_\alpha\\
    [L_{ab},\bar Q_\alphabar] &= \half \Gamma_{ab} \cdot \bar Q_\alphabar\\
    [Q_\alpha, Q_\beta] &= \Gamma^\mu_{\alpha\beta} P_\mu\\
    [\bar Q_\alphabar, \bar Q_\betabar] &= \Gamma^\mu_{\alphabar\betabar} P_\mu~,
  \end{aligned}
\end{equation}
where, as usual,
\begin{equation}
  \Gamma^\mu_{\alpha\beta} := \left<\beps_\alpha, \Gamma^\mu 
    \beps_\beta\right>
  \qquad\text{and}\qquad
  \Gamma^\mu_{\alphabar\betabar} := \left<\bar\beps_\alphabar, \Gamma^\mu 
    \bar\beps_\betabar\right>~.
\end{equation}

The ideal $I<\fk$ is spanned by $P_\mu$, $Q_\alpha$ and $\bar
Q_\alphabar$, whereas the semisimple factor $\fs = \fso(W) \oplus
\fso(W^\perp)$ is spanned by $L_{\mu\nu}$ and $L_{ab}$.  Letting
$P^\mu$, $Q^\alpha$ and $\bar Q^\alphabar$ be the canonical dual basis
for $I^*$, the relations defining the differential in the deformation
complex $C^\bullet := C^\bullet(I;\fk)^{\fs}$ are the following:
\begin{equation}
  \begin{aligned}[m]
    d P^\mu &= \half \Gamma^\mu_{\alpha\beta} Q^\alpha \wedge Q^\beta
    + \half \Gamma^\mu_{\alphabar\betabar} \bar Q^\alphabar \wedge
    \bar Q^\betabar\\
    d Q^\alpha &= 0\\
    d \bar Q^\alphabar &= 0\\
    d P_\mu &= 0\\
    d Q_\alpha &= - \Gamma^\mu_{\alpha\beta} Q^\beta \otimes  P_\mu\\
    d \bar Q_\alphabar &= - \Gamma^\mu_{\alphabar\betabar} \bar
    Q^\betabar \otimes  P_\mu\\
    d L_{\mu\nu} &=  \eta_{\mu\rho} P^\rho \otimes P_\nu -
    \eta_{\nu\rho} P^\rho \otimes P_\mu + \half Q^\alpha \otimes
    \Gamma_{\mu\nu} \cdot Q_\alpha + \half \bar Q^\alphabar \otimes
    \Gamma_{\mu\nu}  \cdot \bar Q_\alphabar\\
    d L_{ab} &= \half Q^\alpha \otimes \Gamma_{ab} \cdot Q_\alpha
    + \half \bar Q^\alphabar \otimes \Gamma_{ab}  \cdot \bar
    Q_\alphabar~.
  \end{aligned}
\end{equation}

There are no $0$-cochains, whereas the space of $1$-cochains is
$3$-dimensional, spanned by $P^\mu \otimes P_\mu$, $Q^\alpha \otimes
Q_\alpha$ and $\bar Q^\alphabar \otimes \bar Q_\alphabar$,
corresponding to the identity maps $W \to W$ and $\Delta_{\NS5}\cap
\Delta_\pm \to \Delta_{\NS5} \cap \Delta_\pm$.  The space of
$2$-cochains is $5$-dimensional, spanned by $P^\mu\wedge P^\nu\otimes
L_{\mu\nu}$, $P^\mu \wedge Q^\alpha \otimes \Gamma_\mu \cdot
Q_\alpha$, $P^\mu \wedge \bar Q^\alphabar \otimes \Gamma_\mu \cdot
\bar Q_\alphabar$, $Q^\alpha \wedge Q^\beta \otimes
\Gamma^\mu_{\alpha\beta} P_\mu$ and $\bar Q^\alphabar \wedge \bar
Q^\betabar \otimes \Gamma^\mu_{\alphabar\betabar} P_\mu$.  Notice that
$\Gamma^{\mu\nu}_{\alpha\betabar} = 0 = \Gamma^{ab}_{\alpha\betabar}$,
whence there are no cochains of the form $Q\bar Q\otimes L$.

Computing the differential $d:C^1 \to C^2$, we see that $H^1(\fk;\fk)
\cong \RR$, with representative cocycle $2 P^\mu
\otimes P_\mu - Q^\alpha \otimes Q_\alpha - \bar Q^\alphabar \otimes
\bar Q_\alphabar$.  This implies that $\dim B^2 = 2$, spanned by $d
(Q^\alpha \otimes Q_\alpha)$ and $d(\bar Q^\alpha \otimes \bar
Q_\alphabar)$, say.  Therefore any cohomology in dimension $2$ must be
represented by a cocycle of the form
\begin{equation}
  \Theta = a_1 P^\mu \wedge P^\nu \otimes L_{\mu\nu} + 
           a_2 P^\mu \wedge Q^\alpha \otimes \Gamma_\mu \cdot Q_\alpha + 
           a_3 P^\mu \wedge Q^\alphabar \otimes \Gamma_\mu \cdot
           Q_\alphabar~.
\end{equation}
As usual, the cocycle condition implies $a_1=0$ because that term is
the only one whose differential involves $L_{\mu\nu}$.  The
differential of the remaining terms have the form $PQ\bar Q\otimes P$,
$QQQ\otimes\bar Q$, $\bar Q\bar Q Q \otimes \bar Q$, $\bar Q\bar
Q\bar Q\otimes Q$ and $QQ \bar Q \otimes Q$.  The $PQ\bar Q\otimes P$
terms are
\begin{multline}
  - P^\mu \wedge Q^\alpha \wedge \bar Q^\betabar \otimes P_\nu \left(
    a_1 \left(\Gamma_\mu\right)^\gammabar{}_\alpha
    \Gamma^\nu_{\gammabar\betabar} + a_2 \left(\Gamma_\mu\right)^\gamma{}_\betabar
    \Gamma^\nu_{\gamma\alpha} \right)\\
  =  P^\mu \wedge Q^\alpha \wedge  \bar Q^\betabar \otimes P_\nu \left( a_1 \left(\Gamma_\mu
      \Gamma^\nu\right)_{\alpha\betabar} + a_2 \left(\Gamma_\mu
      \Gamma^\nu\right)_{\betabar\alpha} \right)~,
\end{multline}
which vanishes because
\begin{equation}
  \begin{split}
    \left(\Gamma_\mu \Gamma^\nu\right)_{\alpha\betabar} &=
    \langle\beps_\alpha, \Gamma_\mu \Gamma^\nu
      \bar\beps_\betabar\rangle \\
    &= \langle\bnu_W  \beps_\alpha, \Gamma_\mu \Gamma^\nu
      \bar\beps_\betabar\rangle \\
    &= - \langle\beps_\alpha, \bnu_W  \Gamma_\mu \Gamma^\nu
      \bar\beps_\betabar\rangle \\
    &= - \langle\beps_\alpha, \Gamma_\mu \Gamma^\nu
      \bnu_W  \bar\beps_\betabar\rangle \\
    &= - \langle\beps_\alpha, \Gamma_\mu \Gamma^\nu
    \bar\beps_\betabar\rangle\\
    &= - \left(\Gamma_\mu \Gamma^\nu\right)_{\alpha\betabar}~,
  \end{split}
\end{equation}
whence $\left(\Gamma_\mu \Gamma^\nu\right)_{\alpha\betabar}=0$
and, similarly, $\left(\Gamma_\mu
  \Gamma^\nu\right)_{\betabar\alpha}=0$.  The $QQQ\otimes\bar Q$ term
also vanishes by polarising identity \eqref{eq:DiracIIA} and the fact
that $\Gamma^a_{\alpha\beta} = 0$.  By a similar argument we see that
the $\bar Q\bar Q\bar Q\otimes Q$ term also vanishes.  The remaining
terms would vanish if and only if
\begin{equation}
  \left<\varepsilon_\pm, \Gamma^\mu  \varepsilon_\pm\right> \Gamma_\mu
  \varepsilon_\mp \stackrel{?}{=} 0 \qquad \forall \varepsilon_\pm \in
  \Delta_{\NS5} \cap \Delta_\pm~.
\end{equation}
It is not hard to show that this is false, proving the rigidity of the
NS5-brane superalgebra.

\subsection{A deformation of the D6-brane superalgebra}
\label{sec:D6}

The D6-brane Killing superalgebra $\fk$ is the subsuperalgebra of the
IIA Poincaré superalgebra associated to a split $V = W \oplus
W^\perp$, with $W$ a $7$-dimensional lorentzian subspace corresponding
to the brane worldvolume.  This means that $\fk_0 = \fso(W) \oplus W
\oplus \fso(W^\perp)$ and $\fk_1$ is isomorphic to the subspace
$\Delta_{\D6} \subset \Delta_+ \oplus \Delta_-$ defined as the graph
of the linear map $\bnu_W : \Delta_+ \to \Delta_-$ corresponding to
the volume element of $W$, which is symmetric relative to the spinor
inner product and satisfies $\bnu_W^2 = + \1$.  Let $\be_\mu$  and
$\be_a$ span $W$ and $W^\perp$, respectively and let $P_\mu$,
$L_{\mu\nu}$ and $L_{ab}$ be the corresponding basis for $\fk_0$.  If
$\beps_\alpha$ is a basis for $\Delta_+$, then $\bpsi_\alpha :=
\tfrac1{\sqrt{2}} \begin{pmatrix}\beps_\alpha\\ \bnu_W 
  \beps_\alpha\end{pmatrix}$ span $\Delta_{\D6}$.  The corresponding
basis for $\fk_1$ is $Q_\alpha$.  The nonzero Lie brackets of $\fk$
are given, in addition to those of $\fk_0$, by
\begin{equation}
  \label{eq:IIAD6KSA}
  \begin{aligned}[m]
    [L_{\mu\nu},Q_\alpha] &= \half \Gamma_{\mu\nu} \cdot Q_\alpha\\
    [L_{ab},Q_\alpha] &= \half \Gamma_{ab} \cdot Q_\alpha\\
    [Q_\alpha, Q_\beta] &= \Gamma^\mu_{\alpha\beta} P_\mu~,
  \end{aligned}
\end{equation}
where
\begin{equation}
  \Gamma^\mu_{\alpha\beta} := \langle\bpsi_\alpha, \Gamma^\mu 
    \bpsi_\beta\rangle = \left<\beps_\alpha, \Gamma^\mu 
    \beps_\beta\right>~.
\end{equation}

The ideal $I<\fk$ is spanned by $P_\mu$ and $Q_\alpha$, whereas the
semisimple factor $\fs = \fso(W) \oplus \fso(W^\perp)$ is spanned by
$L_{\mu\nu}$ and $L_{ab}$.  The canonical dual basis for $I^*$ is
$P^\mu$ and $Q^\alpha$, relative to which the differential in the
deformation complex $C^\bullet := C^\bullet(I;\fk)^{\fs}$ is defined
by the following relations
\begin{equation}
  \begin{aligned}[m]
    d P^\mu &= \half \Gamma^\mu_{\alpha\beta} Q^\alpha \wedge Q^\beta\\
    d Q^\alpha &= 0\\
    d P_\mu &= 0\\
    d Q_\alpha &= - \Gamma^\mu_{\alpha\beta} Q^\beta \otimes  P_\mu\\
    d L_{\mu\nu} &=  \eta_{\mu\rho} P^\rho \otimes P_\nu -
    \eta_{\nu\rho} P^\rho \otimes P_\mu + \half Q^\alpha \otimes
    \Gamma_{\mu\nu} \cdot Q_\alpha\\
    d L_{ab} &= \half Q^\alpha \otimes \Gamma_{ab} \cdot Q_\alpha~.
  \end{aligned}
\end{equation}

There are no $0$-cochains since $\fk^{\fs} = 0$.  The space of
$1$-cochains is two-dimensional, spanned by $P^\mu \otimes P_\mu$ and
$Q^\alpha \otimes Q_\alpha$, corresponding to the identity maps $W \to
W$ and $\Delta_{\D6} \to \Delta_{\D6}$.  The space of
$2$-cochains is $5$-dimensional and is spanned by $P^\mu\wedge
P^\nu\otimes L_{\mu\nu}$, $P^\mu \wedge Q^\alpha \otimes
\Gamma_\mu \cdot Q_\alpha$, $Q^\alpha \wedge Q^\beta \otimes
\Gamma^\mu_{\alpha\beta} P_\mu$, $Q^\alpha \wedge Q^\beta \otimes
\left(\Gamma^{\mu\nu} \bnu_W\right)_{\alpha\beta} L_{\mu\nu}$ and
$Q^\alpha \wedge Q^\beta \otimes \left(\Gamma^{ab}
  \bnu_W\right)_{\alpha\beta} L_{ab}$.

Computing the differential $d:C^1 \to C^2$, we find that $H^1(\fk;\fk)
\cong \RR$, with representative cocycle $2 P^\mu \otimes P_\mu-
Q^\alpha \otimes Q_\alpha$, which means that $\dim B^2 = 1$, spanned
by $d(Q^\alpha\otimes Q_\alpha)$, say.  Computing the differential
$d:C^2 \to C^3$ and employing the usual arguments, we see that any
nontrivial cocycle must be of the form
\begin{equation}
  \Theta = a_1 P^\mu \wedge Q^\alpha \otimes \Gamma_\mu \cdot Q_\alpha
  + a_2 Q^\alpha \wedge Q^\beta \otimes \left(\Gamma^{\mu\nu}
    \bnu_W\right)_{\alpha\beta} L_{\mu\nu} + a_3 Q^\alpha \wedge
  Q^\beta \otimes (\Gamma^{ab} \bnu_W)_{\alpha\beta} L_{ab}~.
\end{equation}
Computing its differential we find two types of terms which must
vanish separately for $\Theta$ to be a cocycle.  The first term takes
the form
\begin{equation}
  P^\mu \wedge Q^\alpha \wedge Q^\beta \otimes P_\nu \left( - a_1
    (\Gamma_\mu \bnu_W)^\gamma{}_\alpha \Gamma^\nu_{\gamma\beta} + 2
    a_2 (\Gamma_\mu{}^\nu \bnu_W)_{\alpha\beta} \right)~.
\end{equation}
Using that
\begin{equation}
  \begin{split}
    (\Gamma_\mu \bnu_W)^\gamma{}_\alpha \Gamma^\nu_{\gamma\beta} &=
    \langle \Gamma_\mu \bnu_W \beps_\alpha, \Gamma^\nu
    \beps_\beta\rangle \\
    &= - \langle \bnu_W \beps_\alpha, \Gamma_\mu \Gamma^\nu
    \beps_\beta \rangle \\
    &= - \langle \beps_\alpha, \Gamma_\mu \Gamma^\nu
    \bnu_W \beps_\beta \rangle \\
    &= - (\Gamma_\mu{}^\nu \bnu_W)_{\alpha\beta} - \delta_\mu^\nu
    \langle \beps_\alpha, \bnu_W \beps_\beta \rangle~,
  \end{split}
\end{equation}
and that the second term is skewsymmetric in
$\alpha\leftrightarrow\beta$, the above term in $d\Theta$ becomes
\begin{equation}
  ( a_1 + 2 a_2 ) (\Gamma_\mu{}^\nu \bnu_W)_{\alpha\beta} P^\mu \wedge
  Q^\alpha \wedge Q^\beta \otimes P_\nu~,
\end{equation}
whence $d\Theta = 0$ forces $a_1 = -2 a_2$.  The second type of term
in $d\Theta$ is given by
\begin{equation}
  \half Q^\alpha \wedge Q^\beta \wedge Q^\gamma \otimes Q_\delta
  \left( a_1 \Gamma^\mu_{\alpha\beta} \left(\Gamma_\mu
      \bnu_W\right)^\delta{}_\gamma + a_2
    \left(\Gamma^{\mu\nu} \bnu_W\right)_{\alpha\beta}
    \left(\Gamma_{\mu\nu}\right)^\delta{}_\gamma + a_3 
    (\Gamma^{ab} \bnu_W)_{\alpha\beta}
    \left(\Gamma_{ab}\right)^\delta{}_\gamma\right)~,
\end{equation}
whose vanishing is equivalent, via a polarisation identity, to the
vanishing of
\begin{equation}
  a_1 \left<\varepsilon, \Gamma^\mu  \varepsilon\right> \Gamma_\mu
   \bnu_W  \varepsilon + a_2 \left<\varepsilon,
    \Gamma^{\mu\nu}  \bnu_W  \varepsilon\right>
  \Gamma_{\mu\nu}  \varepsilon + a_3 \langle\varepsilon,
    \Gamma^{ab}  \bnu_W  \varepsilon\rangle
  \Gamma_{ab}  \varepsilon = 0
\end{equation}
for all $\varepsilon \in \Delta_+$.  Using the Clifford identities
\begin{equation}
  \Gamma^{ab} \bnu_W \varepsilon = \epsilon^{abc} \Gamma_c \varepsilon
  \qquad\text{and}\qquad
  \half \epsilon^{abc} \Gamma_{bc} \varepsilon = \Gamma^a \bnu_W \varepsilon
\end{equation}
for $\varepsilon \in \Delta_+$, we see that
\begin{equation}
  \begin{split}
    \langle\varepsilon, \Gamma^{ab} \bnu_W \varepsilon\rangle
    \Gamma_{ab} \varepsilon &= 2 \langle\varepsilon, \Gamma^c \varepsilon\rangle
    \Gamma_c \bnu_W \varepsilon\\
    &= - 2 \bnu_W \langle\varepsilon, \Gamma^c \varepsilon\rangle
    \Gamma_c \varepsilon\\
    &= 2 \bnu_W \langle\varepsilon, \Gamma^\mu \varepsilon\rangle
    \Gamma_\mu \varepsilon\\
    &= 2 \langle\varepsilon, \Gamma^\mu \varepsilon\rangle
    \Gamma_\mu \bnu_W \varepsilon~,    
  \end{split}
\end{equation}
using equation \eqref{eq:DiracIIA}.  Plugging the above into the Fierz identity
\eqref{eq:fierz10dbrane3}, we find
\begin{equation}
  \left<\varepsilon, \Gamma^{\mu\nu}  \bnu_W  \varepsilon\right>
  \Gamma_{\mu\nu} \varepsilon = -10 \langle\varepsilon, \Gamma^\mu
  \varepsilon\rangle \Gamma_\mu \bnu_W \varepsilon~,
\end{equation}
whence the second part of the cocycle condition $d\Theta=0$ becomes
\begin{equation}
  (a_1 - 10 a_2 + 2 a_3) \left<\varepsilon, \Gamma^\mu
    \varepsilon\right> \Gamma_\mu \bnu_W  \varepsilon = 0
\end{equation}
for all $\varepsilon \in \Delta_+$, which implies $a_1 = 10 a_2 - 2
a_3$.  Putting both conditions together, we find that the nontrivial
cocycle is a multiple of
\begin{equation}
  \Theta = -2 P^\mu \wedge Q^\alpha \otimes \Gamma_\mu \cdot Q_\alpha
  +  Q^\alpha \wedge Q^\beta \otimes \left(\Gamma^{\mu\nu}
    \bnu_W\right)_{\alpha\beta} L_{\mu\nu} + 6 Q^\alpha \wedge
  Q^\beta \otimes (\Gamma^{ab} \bnu_W)_{\alpha\beta} L_{ab}~,
\end{equation}
which shows that $H^2(\fk;\fk) \cong \RR$.

To first order in the deformation parameter $t$, the nonvanishing Lie
brackets corresponding to the above cocycle, in addition to those of
$\fs$ which do not deform, are given by
\begin{equation}
  \begin{aligned}[m]
    [P_\mu, Q_\alpha] &= 2 t \Gamma_\mu \cdot Q_\alpha \\
    [Q_\alpha, Q_\beta] &= \Gamma^\mu_{\alpha\beta} P_\mu - 2 t
    (\Gamma^{\mu\nu}\bnu_W)_{\alpha\beta} L_{\mu\nu} - 12 t 
    (\Gamma^{ab}\bnu_W)_{\alpha\beta} L_{ab}~. 
  \end{aligned}
\end{equation}
There is an obstruction to integrating this deformation at order
$t^2$, which can be overcome by introducing the bracket
\begin{equation}
  [P_\mu, P_\nu] = 16 t^2 L_{\mu\nu}~.
\end{equation}
One can check that the above Lie brackets now satisfy Jacobi for all
$t$.  When $t\neq 0$, one can rescale $P_\mu$ and $Q_\alpha$ in order
to get rid of $t$ and bring the Lie algebra to the following form
\begin{equation}
  \label{eq:IIAD6deformed}
  \begin{aligned}[m]
    [P_\mu, P_\nu] &= L_{\mu\nu}\\
    [P_\mu, Q_\alpha] &= \half \Gamma_\mu \cdot Q_\alpha \\
    [Q_\alpha, Q_\beta] &= \pm \left( \Gamma^\mu_{\alpha\beta} P_\mu -
      \half (\Gamma^{\mu\nu}\bnu_W)_{\alpha\beta} L_{\mu\nu} - 3
      (\Gamma^{ab}\bnu_W)_{\alpha\beta} L_{ab}\right)~.
  \end{aligned}
\end{equation}
The bosonic subalgebra is now $\fso(2,6) \oplus \fso(3)$ and the Lie
superalgebras above are real forms of the simple Lie superalgebra of
type $D(4,1)$ in Kac's classification \cite{KacSuperSketch}.  In fact,
it is isomorphic to $\fosp(6,2|2)$, which is the conformal
superalgebra of six-dimensional Minkowski spacetime, which suggests
that the worldvolume of the D6-brane curves to $\AdS_7$.

This deformation is related via Kaluza--Klein reduction to the similar
deformation of the Kaluza--Klein monopole superalgebra in
equation~(95) of \cite[Section~5.2]{JMFSuperDeform}.

\section{Conclusions}
\label{sec:conclusions}

In this paper we have explored the Lie superalgebra deformations of
the Killing superalgebras of ten-dimensional supergravity
backgrounds.  We have concentrated largely on the flat vacua, which
have been shown to be rigid, and the elementary asymptotically flat
branes.  All have been found to be rigid except for the following:
\begin{itemize}
\item Types I and IIB D1-brane superalgebra, with deformed
  superalgebras given by \eqref{eq:ID1deformed} and
  \eqref{eq:IIBD1deformed}, respectively;
\item Types IIA and IIB fundamental string superalgebras, with deformations
  \eqref{eq:IIAF1deformed} and \eqref{eq:IIBF1deformed}, respectively;
\item Type IIA D2-brane superalgebra, with deformation
  \eqref{eq:IIAD2deformed}, isomorphic to $\fso(2,1)  \oplus \ff(4)$;
\item Type IIA D6-brane superalgebra, with deformation
  \eqref{eq:IIAD6deformed}, isomorphic to $\fosp(6,2|2)$; and
\item Type IIB D7-brane superalgebra, with deformation
  \eqref{eq:IIBD7deformed}.
\end{itemize}
In particular, these two latter deformations are related to the
deformation of the delocalised M2-brane found in Section \ref{sec:DM2}
and that of the Kaluza--Klein monopole background of
eleven-dimensional supergravity in equation~(95) of
\cite[Section~5.2]{JMFSuperDeform}.

These results seem to suggest a geometric origin for the deformations
found in this paper, simply because it would otherwise be difficult to
justify their good behaviour under a geometric process such as
Kaluza--Klein reduction.  If a background has a symmetry superalgebra
$\fg = \fg_0 \oplus \fg_1$ and we consider its Kaluza--Klein reduction
along the one-parameter subgroup generated by some element $X \in
\fg_0$, then the symmetry superalgebra $\fk$ of the quotient is
isomorphic to $\fn/\fh_X$, where $\fh_X < \fg_0$ is the
one-dimensional Lie subalgebra spanned by $X$ and $\fn < \fg$ is the
normaliser of $\fh_X$ in $\fg$.  As a Lie superalgebra, $\fk = \fk_0
\oplus \fk_1$, where $\fk_1$ are those elements of $\fg_1$ which
commute with $X$ and $\fk_0$ is the normaliser of $X$ in $\fg_0$
modulo the span of $X$.  It is easy to check that for the delocalised
M2 brane and the D2, which is its reduction along the delocalised
direction, and for the Kaluza--Klein monopole and the D6, which is its
reduction along the central element, the Killing superalgebras do
indeed behave in the way just stated and \emph{moreover} so do their
deformations.  It is precisely this coherence under Kaluza--Klein
reduction which suggests that the deformations do have a geometric
construction.  Furthermore, as explained in Section \ref{sec:IIAF1},
the deformation of the M2 brane does not induce a deformation of the
fundamental string because the dimension of the superalgebras are
different.  Although it has not been the purpose of this paper to
elucidate the geometric interpretation of the deformations found here
and in \cite{JMFSuperDeform}---this will be reported on in
\cite{FigRitDef}---we nevertheless believe that we have given evidence
that such an interpretation ought to exist.

\section*{Acknowledgments}

We have enjoyed discussions about this topic with Patricia Ritter and
Joan Simón.  Part of this work was done while BV was visiting the
School of Mathematics of the University of Edinburgh, with support
from the Marie Curie Research Training Network Grant
``ForcesUniverse'' (contract no. MRTN-CT-2004-005104) from the
European Community's Sixth Framework Programme.

\appendix

\section{Spinorial conventions}
\label{sec:spin}

We work with a mostly plus metric and with a plus sign in the Clifford
algebra.  Let $V$ denote an eleven-dimensional lorentzian vector space
with signature $(10,1)$ and let $\Cl(V)$ denote its Clifford algebra.
It is well-known that $\Cl(V) \cong \End(\Delta) \oplus
\End(\Delta')$, where the irreducible Clifford modules $\Delta$ and
$\Delta'$ are real and 32-dimensional and are distinguished by the
action of the central element $\bnu_V$ corresponding to the volume
form on $V$.  We will work with $\Delta$, say.  Introduce an
orthonormal frame $\be_0,\be_1,\dots,\be_\ten$ for $V$ and denote the
corresponding elements in $\Cl(V)$ by $\Gamma_0,\dots,\Gamma_\ten$.
Let $W = \left<\be_\ten\right>^\perp$ be the ten-dimensional
lorentzian subspace perpendicular to $\be_\ten$.  Although $\Delta$ is
irreducible under $\fso(V)$, it decomposes under $\fso(W)$ as $\Delta
= \Delta_+ \oplus \Delta_-$, each summand being an eigenspace of
$\Gamma_\ten$.  The invariant symplectic form $\left<-,-\right>$ on
$\Delta$ is such that $\Delta_\pm$ are lagrangian subspaces.  If $\psi
\in \Delta$, then there is an eleven-dimensional Fierz-type identity
which reads
\begin{equation}
  \label{eq:fierz11}
  \half \left<\psi, \Gamma^{MN} \psi\right> \Gamma_{MN} \psi = - 5
  \left<\psi, \Gamma^M \psi\right> \Gamma_M \psi~,
\end{equation}
and a similar identity involving the natural 5-form:
\begin{equation*}
  \tfrac1{5!} \left<\psi, \Gamma^{M_1 \cdots M_5} \psi\right>
  \Gamma_{M_1 \cdots M_5} \psi = - 6 \left<\psi, \Gamma^M \psi\right>
  \Gamma_M \psi~,
\end{equation*}
which is consistent with the Fierz identity
\begin{equation}
  \label{eq:bispinor}
  32 \lambda_\psi = \left<\psi, \Gamma^M \psi\right> \Gamma_M - 
  \half \left<\psi, \Gamma^{MN} \psi\right> \Gamma_{MN} +
  \tfrac1{5!} \left<\psi, \Gamma^{M_1 \cdots M_5} \psi\right>
  \Gamma_{M_1 \cdots M_5}~,
\end{equation}
with $\lambda_\psi \in \End(\Delta)$ the rank-1 endomorphism defined by
$\lambda_\psi (\chi) = \left<\psi,\chi\right>\psi$.

We may reinterpret identity \eqref{eq:fierz11} in ten dimensions
as follows.  Let $\psi = \begin{pmatrix}\varepsilon_+ \\
  \varepsilon_- \end{pmatrix}$, with $\varepsilon_\pm \in \Delta_\pm$.
Using that for any $\varepsilon \in \Delta_\pm$,
\begin{equation}
  \label{eq:dirac}
  \left<\varepsilon, \Gamma^A\varepsilon\right> \Gamma_A \varepsilon =
  0~,
\end{equation}
where $A=0,1,\dots,9$, we may unpack equation \eqref{eq:fierz11} as
\begin{equation}
  \label{eq:fierz10}
  4 \left<\varepsilon_\pm, \Gamma^A \varepsilon_\pm\right> \Gamma_A
  \varepsilon_\mp - 10 \left<\varepsilon_\pm, \varepsilon_\mp\right>
  \varepsilon_\pm + \left<\varepsilon_\pm, \Gamma^{AB}
    \varepsilon_\mp\right> \Gamma_{AB} \varepsilon_\pm = 0~.
\end{equation}

In this paper we will need the restriction of this identity to various
subspaces of $\Delta$ or $\Delta_+$ or of $2\Delta_+$, depending on
the supergravity theory in question.  For type II D-branes, we will be
interested in the graphs of linear maps $\bnu_W: \Delta_+ \to
\Delta_\pm$ inside $\Delta_+ \oplus \Delta_\pm$, with $\bnu_W$ the
volume form of a lorentzian subspace of $V$ associated to a brane
worldvolume.

In the case of IIB D-branes, $W$ is even-dimensional, and hence 
$\bnu_W : \Delta_+ \to \Delta_+$.  The Fierz identity of relevance is
the restriction of equation \eqref{eq:dirac}, which now says
\begin{equation}
  \label{eq:diracW}
  \left<\varepsilon, \Gamma^\mu\varepsilon\right> \Gamma_\mu \varepsilon =
  0~,
\end{equation}
for $\mu=0,\dots,\dim W$.

In the case of IIA D-branes, $W$ is odd-dimensional, and hence $\bnu_W
: \Delta_+ \to \Delta_-$ and its graph consists of spinors with
$\varepsilon_- = \bnu_W \varepsilon_+$.  In this case, the two
identities \eqref{eq:fierz10} become equivalent to this one:
\begin{equation}
  \label{eq:fierz10brane}
  4 \left<\varepsilon, \Gamma^A \varepsilon\right> \Gamma_A
  \bnu_W \varepsilon - 10 \left<\varepsilon, \bnu_W
    \varepsilon\right> \varepsilon + \left<\varepsilon,
    \Gamma^{AB} \bnu_W \varepsilon\right> \Gamma_{AB} \varepsilon = 0~,
\end{equation}
for all $\varepsilon \in \Delta_+$.  We can refine this identity
further.  Let $\sigma_W := - (-1)^{p/2}$, where $\dim W = p + 1$ with
$p$ even.  The it follows that
$\left<\bnu_W\varepsilon_1,\varepsilon_2\right> = \sigma_W
\left<\varepsilon_1, \bnu_W \varepsilon_2\right>$ and similarly that
$\bnu_W^2 = \sigma_W \1$.  Letting $\be_\mu$ and $\be_a$ denote
orthonormal frames for $W$ and $W^\perp$, respectively, we have that
\begin{equation*}
  \bnu_W \Gamma^\mu = \Gamma^\mu \bnu_W \qquad\text{whereas}\qquad
  \bnu_W \Gamma^a = -\Gamma^a \bnu_W ~.
\end{equation*}
Using this it is not hard to show that if $\sigma_W = -1$, so that
$\dim W \equiv 1 \pmod 4$, then $\left<\varepsilon,
  \Gamma^{\mu\nu}\bnu_W \varepsilon\right> = 0$ and 
$\left<\varepsilon,  \Gamma^{ab}\bnu_W \varepsilon\right> = 0$~,
whence the identity \eqref{eq:fierz10brane} becomes
\begin{equation}
  \label{eq:fierz10dbrane1}
  4 \left<\varepsilon, \Gamma^\mu \varepsilon\right> \Gamma_\mu
  \varepsilon + 5 \left<\varepsilon, \bnu_W
    \varepsilon\right> \bnu_W \varepsilon + \left<\varepsilon,
    \Gamma^{\mu a} \bnu_W \varepsilon\right> \Gamma_{\mu a} \bnu_W
  \varepsilon = 0~.  
\end{equation}
On the other hand, if $\sigma_W = 1$, so that
$\dim W \equiv 3 \pmod 4$, then $\left<\varepsilon,
  \Gamma^{\mu a}\bnu_W \varepsilon\right> = 0$ and 
$\left<\varepsilon,  \bnu_W \varepsilon\right> = 0$~,
whence the identity \eqref{eq:fierz10brane} becomes
\begin{equation}
  \label{eq:fierz10dbrane3}
  8 \left<\varepsilon, \Gamma^\mu \varepsilon\right> \Gamma_\mu
  \varepsilon +  \left<\varepsilon, \Gamma^{\mu\nu}\bnu_W
    \varepsilon\right> \Gamma_{\mu\nu} \bnu_W \varepsilon + 
  \left<\varepsilon, \Gamma^{ab}\bnu_W
    \varepsilon\right> \Gamma_{ab} \bnu_W \varepsilon = 0~.
\end{equation}

For branes which are not D-branes, e.g., the fundamental string and
the NS5-brane, the subspace of Killing spinors is not naturally a
graph.  Any Fierz identities used in those calculations will be
recalled as needed.

\bibliographystyle{utphys}
\bibliography{AdS,AdS3,ESYM,Sugra,Geometry,Algebra}

\end{document}